\documentclass[twocolumn,10pt,aps,prd]{revtex4-2}
\usepackage{hyperref,graphicx,amsfonts,amsmath,amssymb}
\usepackage[latin1]{inputenc}
\usepackage{multirow}
\usepackage{subcaption}
\usepackage{diagbox}
\usepackage{titlesec}
\usepackage{soul}
\usepackage{xcolor}
\sethlcolor{lime}
\newcommand{\tn}{\textnormal}
\newcommand{\GeV}{\tn{GeV}}
\newcommand{\TeV}{\tn{TeV}}
\newcommand{\PartTitle}[1]{
	\begin{center}
		\vspace*{0.5cm}
		{\Large\bfseries #1}
		\vspace*{0.5cm}
	\end{center}
}
\def\beq{\begin{equation}}
\def\eeq{\end{equation}}
\def\beqa{\begin{eqnarray}}
\def\eeqa{\end{eqnarray}}
\def\sba{s_{\beta-\alpha}}
\def\cba{c_{\beta-\alpha}}

\def\sb{s_{\beta}}
\def\cb{c_{\beta}}
\def\tb{t_{\beta}}
\def\ctb{ct_{\beta}}
\def\calm{\mathcal{M}}

\def\call{\mathcal{L}}
\def\hpm{{H^\pm}}

\def\wpm{{W^\pm}}
\def\hp{{H^+}}

\def\mh{m_{h_{125}}}
\def\mH{m_{H}}
\def\mhpm{m_{\hpm}}
\def\ma{{m_A}}

\def\m12{m_{12}}
\def\h125{h_{125}}
\begin{document}
\title{Observability of 2HDM charged Higgs boson in a misaligned scenario}

\author{Majid Hashemi}
\email{hashemi$_$mj@shirazu.ac.ir}
\author{Omid Seify}
\email{omidseify@hafez.shirazu.ac.ir}
\affiliation{Department of Physics, College of Science, Shiraz University, Shiraz, 71946-84795, Iran}

\begin{abstract}
The large hadron collider precision measurement of $\h125$ properties suggests that a misaligned two Higgs doublet model type I is preferred among other types of the model. In this analysis, a lepton collider potential for charged Higgs boson observation is evaluated and compared with HL-LHC expectations. We focus on $\hpm \to \wpm \h125$ decay which is exclusively activated in misaligned scenario. An event analysis including fast detector simulation is performed and it is shown that some parts of the parameter space which are out of HL-LHC reach can be probed by a high energy lepton collider with signal significance exceeding $2\sigma$ and $5\sigma$ for exclusion or discovery. The region with highest sensitivity is in the middle range charged Higgs boson mass, i.e., $350<\mhpm<450$ GeV and $\tan\beta \gtrsim4$ at non-zero $\cos(\beta-\alpha)$ values.    
\end{abstract}

\maketitle

\section{Introduction}
More than a decade has passed since the observation of a new resonance around 125 GeV by the two experiments CMS and ATLAS at the Large Hadron Collider (LHC) \cite{HiggsCMS,HiggsATLAS}. The observed particle is consistent with standard Model (SM) prediction through the Higgs Mechanism \cite{Higgs1,Higgs2,Englert1,kibble1,Higgs3,kibble2} and its observation was awarded the Nobel prize in Physics in 2013.

Soon after the discovery, both collaborations have focused on measuring properties of the Higgs boson in terms of production cross section and decay rates \cite{LHC1,LHC5,LHC6,LHC7,LHC8,CMStautau}, its couplings with other particles \cite{tH}, self-coupling \cite{ATLASHPair,CMSHPair}, spin and parity \cite{LHC4}, rare decays \cite{H2mu,H2llph}, CP structure \cite{LHC2,LHC3,phi}, etc. 

Despite the overall agreement between the measured properties and the SM prediction, some deviations are observed which are expressed in terms of the signal strength ($\mu$) and the coupling modifier ($\kappa$) in different production and decay channels \cite{portrait}.  

The CMS collaboration with $\h125\to \tau\tau$ production measurement has quoted the signal strength $\mu=0.85^{+0.12}_{-0.11}$ in \cite{CMStautau}. The Higgs boson coupling with the top quark has been analyzed through $t\bar{t}\h125$ and $t\h125$ production and $\kappa_t$ (the measured coupling normalized to SM) is within $-0.9<\kappa_t<-0.7$ or $0.7<\kappa_t<1.1$ \cite{tH}. The signal strength in these channels has been measured to be $0.92\pm0.19(\tn{stat})_{-0.13}^{+0.17}(\tn{syst})$ and $5.7\pm2.7(\tn{stat})\pm3.0(\tn{syst})$ respectively. The Higgs boson pair production has also been among the search channels due to the possibility of Higgs boson self-coupling studies through $\lambda$ parameter in the Higgs potential. The measured coupling modifier $\kappa_{\lambda}$ is found to be $-1.5<\kappa_{\lambda}<6.7$ (ATLAS)\cite{ATLASHPair} and $-3.3<\kappa_{\lambda}<8.5$ (CMS)\cite{CMSHPair}. The rare Higgs boson decays have also been analyzed including $\h125\to \mu\mu$ (CMS) with $\mu=1.19^{+0.4}_{-0.39}(\tn{stat})^{+0.15}_{-0.14}(\tn{syst})$ \cite{H2mu} and $\h125\to \ell\ell\gamma$ (ATLAS) with $\mu=1.5\pm0.5$ \cite{H2llph}. The CP structure of the observed boson has been measured in terms of the CP-even and CP-odd mixing angle $\phi$ and the result is found to be $\phi=4 \pm 36^\circ$ in \cite{phi}. 

These results are examples of analyses performed by the two LHC collaborations showing overall compatibility with SM despite observed deviations in some channels. While more data is expected to reduce statistical uncertainties, BSM interpretation of the results is also among the possible choices if it turns out that the current deviations from SM are not statistical fluctuations but rather meaningful observations. In such direction, one of the most attractive scenarios is the extended Higgs sector which is characterized by a rich family of Higgs bosons with non-SM couplings with matter. 

Models with extended Higgs sectors are not only able to explain Higgs-fermion or Higgs-gauge coupling deviations from SM but are also among the main candidates which provide solutions to open problems such as the hierarchy problem. The theory of supersymmetry is built upon two Higgs doublets to provide a natural solution for the hierarchy in the Higgs sector i.e., the divergence of the Higgs boson mass radiative corrections \cite{HTPH,habersusy}.  

Some scenarios of extended Higgs sector include a scalar doublet with no interaction with SM particles (inert doublet) and suggest dark matter candidates \cite{IDM0,IDM1,IDM2,IDM3,IDM4,IDM5,IDMCS}. The theory of supersymmetry has also its lightest super-symmetric particle (LSP) as the dark matter candidate in scenarios with R-parity conservation which prevent LSP from decaying to SM particles \cite{SUSYDM}.

Some Models of CP violation through baryon asymmetry at the electroweak scale \cite{CPViolation} or the neutrino mass \cite{NuMass} require extended Higgs sectors.
 
The two Higgs doublet model (2HDM) is one of the main scenarios of BSM with extended Higgs sector \cite{2hdm1}. 
The Higgs sector of 2HDM incorporates two Higgs doublets resulting in five physical Higgs bosons \cite{2hdm1,2hdm2,2hdm3}, i.e., three neutral bosons ($\h125,~H,~A$) and two charged bosons ($H^{\pm}$) \cite{2hdm_HiggsSector1,2hdm_HiggsSector2}. 

There are four types of CP conserving 2HDM without FCNC which are characterized in terms of Higgs-fermion couplings \cite{Barger_2hdmTypes,2hdm_TheoryPheno}. The Minimal Super-symmetric SM (MSSM) uses 2HDM type II as the Higgs sector \cite{MSSM1,MSSM2,MSSM3,MSSM4}. 

The lightest neutral Higgs boson in 2HDM can be aligned at tree level with SM Higgs boson ($h_{SM}$) in terms of the Higgs-fermion couplings. This is the so called alignment limit which respects the LHC observation at 125 GeV with SM-like properties \cite{align1,align2,align3}. The alignment occurs at the decoupling limit, i.e., when the mass difference between the light and heavy Higgs bosons is large \cite{decoupling,pos5}. However, it can also be realized naturally without decoupling or fine tuning \cite{m12}.

While many analyses have focused on the standard alignment scenario, this work is going to address the misaligned 2HDM through a charged Higgs search channel with $H^{\pm} \to W^{\pm}\h125$. 

The next sections, include theoretical background, current theoretical and experimental constraints on the charged Higgs boson, collider choice, signal identification, software setup and details of the analysis and results followed by conclusions.

\section{Theoretical Framework}	
\label{thfr}
A natural extension of the SM Higgs sector leads to the 2HDM Lagrangian which contains two complex scalar doublets $\Phi_1$ and $\Phi_2$ in the kinetic term and the potential: 
\begin{eqnarray}\label{lagrangian-2HDM}
	\mathcal L_{\Phi}^{\textnormal{2HDM}}&=& \sum_{i=1,2}(D_{\mu}\Phi_{i})^{\dagger}(D^{\mu}\Phi_{i})-\mathcal V^{\textnormal{2HDM}}_{\Phi}.
\end{eqnarray}
The covariant derivative $D_\mu$ acting on $\Phi_i$ results in Higgs-gauge interactions which are determined before the four types are introduced through Higgs-fermion interactions. The two doublets can be written as \cite{2hdm_TheoryPheno}:
\begin{equation}\label{phi2HDM}
	\Phi_{i}=\binom{\phi_{i}^{+}}{(v_{i}+\rho_i+i\eta_i)/\sqrt{2}},~ i=1, 2 
\end{equation}
where $v_{i}$ ($v_1=v c_\beta,~v_2=v s_\beta$) are vacuum expectation values (vevs) of the two doublets under the norm condition $v=\sqrt{v_1^2+v_2^2}=246 ~\textnormal{GeV}$. The usual abbreviations, $\sin\beta=s_\beta,~\cos\beta=c_\beta,~\tan\beta=\tb,~\cos(\beta-\alpha)=\cba$, etc. are used throughout the paper. 

Introducing two mixing angles $\alpha$ and $\beta$, the physical Higgs bosons are obtained through orthogonal combinations  
\begin{align}
	\binom{H}{\h125}&=\left (\begin{matrix}c_\alpha & s_\alpha\\-s_\alpha &c_\alpha\end{matrix}\right )\binom{\rho_1}{\rho_2} \\ 
	A&=c_\beta\eta_2-s_\beta\eta_1  \\ 
	H^{\pm}&=c_\beta\phi^{\pm}_2-s_\beta\phi^{\pm}_1.
	\label{mixing}
\end{align}
The charged Higgs boson is obtained through the scalar fields $\phi^{\pm}_{i}$ with no vev to avoid the charged vacuum expectation value.

The SM Higgs boson would be realized as a combination of CP-even scalars:
\begin{equation}\label{hsm}
	h_{\textnormal{SM}}=\rho_1 c_\beta + \rho_2 s_\beta = \h125~ \sba-H~ \cba.
\end{equation}
Therefore if $s_{\beta-\alpha}=1$, i.e., when $\beta-\alpha=\pi/2$, the lightest 2HDM scalar is SM-like \cite{tanbsignificance}.
 
The Higgs potential containing the mass terms and self-interactions in CP-conserving form is written as follows:
\begin{align}
	\mathcal{V}^{\textnormal{2HDM}}_{\Phi} \nonumber &= m_{11}^2\Phi_1^\dagger\Phi_1+m_{22}^2\Phi_2^\dagger\Phi_2-m_{12}^2\left(\Phi_1^\dagger\Phi_2+\Phi_2^\dagger\Phi_1\right)\\
	\nonumber &+\frac{1}{2}\lambda_1\left(\Phi_1^\dagger\Phi_1\right)^2+\frac{1}{2}\lambda_2\left(\Phi_2^\dagger\Phi_2\right)^2\\ \nonumber
	\nonumber &+\lambda_3\left(\Phi_1^\dagger\Phi_1\right)\left(\Phi_2^\dagger\Phi_2\right)+\lambda_4\left(\Phi_1^\dagger\Phi_2\right)\left(\Phi_2^\dagger\Phi_1\right)\\ \nonumber
	\nonumber &+\frac{1}{2}\lambda_5\left[\left(\Phi_1^\dagger\Phi_2\right)^2+\left(\Phi_2^\dagger\Phi_1\right)^2\right].\\ 
\label{potential}
\end{align}
Given a set of values for $\beta,~m_{12}^2$ and $\lambda_i$, the Higgs boson masses are obtained from the potential leading to the following values for the CP-odd and charged Higgs bosons:
\begin{align}
m_A^2& \nonumber =\frac{m_{12}^2}{s_\beta c_\beta}-\lambda_5v^2\\ \nonumber
m_{H^{\pm}}^2&=m_A^2+\frac{1}{2}v^2(\lambda_5-\lambda_4).\\
\end{align}
For CP-even Higgs bosons, ($h,~H$), the squared mass matrix is
\beq
\mathcal{M}^2 \equiv  
\left( \begin{array}{cc}
	m_A^2\sb^2+(\lambda_1\cb^2+\lambda_5\sb^2)v^2
	&\sb\cb((\lambda_3+\lambda_4)v^2-m_A^2) \\
	\sb\cb((\lambda_3+\lambda_4)v^2-m_A^2) 
	&m_A^2\cb^2+(\lambda_2\sb^2+\lambda_5\cb^2)v^2
\end{array}\right), 
\label{curlybdef}
\eeq
whose eigenvalues, expressed in terms of the four elements $\calm^2_{ij}$, are
\beq
m^2_{H,\h125}=\frac{1}{2}\left[{\cal M}_{11}^2+{\cal M}_{22}^2
\pm \sqrt{({\cal M}_{11}^2-{\cal M}_{22}^2)^2 +4({\cal M}_{12}^2)^2}
\ \right].
\label{higgsmasses}
\eeq
The Higgs-fermion couplings are introduced through a Yukawa Lagrangian 
\begin{align}
	\mathcal{L}_{Y} \nonumber
	&=\sum_{f=u,d,\ell}\frac{m_f}{v}\Big(\xi_{h_{125}}^{f}\bar{f}fh_{125} + \xi_{H}^{f}\bar{f}fH - i\xi_{A}^{f}\bar{f}\gamma_5fA \Big)\\ \nonumber
	&+\frac{1}{\sqrt{2}v} H^{+} [ V_{ij} m_{u_i} \xi^u_A \bar{u}_i(1 - \gamma^5) d_j \\ \nonumber
	&+ V_{ij} m_{d_j} \xi^d_A \bar{u}_i(1 + \gamma^5) d_j \\ \nonumber
	&+ m_l \xi^l_A \bar{\nu}_i (1 +\gamma^5)l_i ] + h.c.\\
	\label{lag}
\end{align}
where $V_{ij}$ are CKM matrix elements and coupling factors $\xi$ are in general functions of both $\alpha$ and $\beta$ parameters as given in Tab. \ref{couplings}:  
\begin{table}[h]
	\begin{center}
		\begin{tabular}{|c|c|c|c|c|}  \hline
			2HDM type & I & II & X & Y\\
			\hline
			$\xi^u_{h_{125}} $ & $c_\alpha/s_\beta$ & $c_\alpha/s_\beta$ & $c_\alpha/s_\beta$ & $c_\alpha/s_\beta$ \\
			\hline
			$\xi^d_{h_{125}} $ & $c_\alpha/s_\beta$ & $-s_\alpha/c_\beta$ & $c_\alpha/s_\beta$ & $-s_\alpha/c_\beta$ \\
			\hline
			$\xi^\ell_{h_{125}} $ & $c_\alpha/s_\beta$ & $-s_\alpha/c_\beta$ & $-s_\alpha/c_\beta$ & $c_\alpha/s_\beta$ \\
			\hline
			$\xi^u_H $ & $s_\alpha/s_\beta$ & $s_\alpha/s_\beta$ & $s_\alpha/s_\beta$ & $s_\alpha/s_\beta$ \\
			\hline
			$\xi^d_H $ & $s_\alpha/s_\beta$ & $c_\alpha/c_\beta$ & $s_\alpha/s_\beta$ & $c_\alpha/c_\beta$ \\
			\hline
			$\xi^\ell_H $ & $s_\alpha/s_\beta$ & $c_\alpha/c_\beta$ & $c_\alpha/c_\beta$ & $s_\alpha/s_\beta$ \\
			\hline
			$\xi^u_A $ & $ct_\beta$ & $ct_\beta$ & $ct_\beta$ & $ct_\beta$ \\
			\hline
			$\xi^d_A $ & $- ct_\beta$ & $t_\beta$ & $-ct_\beta$ & $t_\beta$ \\
			\hline
			$\xi^\ell_A $ & $- ct_\beta$ & $t_\beta$ & $t_\beta$ & $-ct_\beta$ \\
			\hline
		\end{tabular}
	\end{center}
	\caption{Higgs-fermion Yukawa couplings in different types of 2HDM. Type X and Y are also known as lepton-specific and flipped. }
	\label{couplings}
\end{table}

Any deviation of these coupling factors from unity results in observation of non-SM couplings. Using trigonometric relations
\begin{align}
	s_\alpha/s_\beta \nonumber&~=~ c_{\beta-\alpha}-ct_\beta~ s_{\beta-\alpha}\\ \nonumber
	c_\alpha/c_\beta &~=~ c_{\beta-\alpha} + t_\beta~ s_{\beta-\alpha}\\\nonumber
	-s_\alpha/c_\beta &~=~ s_{\beta-\alpha}-t_\beta ~c_{\beta-\alpha}\\\nonumber
	c_\alpha/s_\beta &~=~ s_{\beta-\alpha} + ct_\beta~ c_{\beta-\alpha}\\
	\label{trig}
\end{align}
the heavy neutral CP-even Higgs boson receives simple $\beta$-dependent coupling with fermions at the alignment limit defined through $\sba \to 1$. The light neutral Higgs couplings with fermion $f$ in such limit tend to their SM form of $m_f/v$.  However, the pseudo-scalar Higgs and charged Higgs-fermion couplings are always in the form of $\tan\beta$ or $\cot\beta$.

The charged Higgs decay to gauge bosons (which is the topic of this work) takes the form $H^{\pm} \to W^{\pm}\phi$ with $\phi=h_{125}/H/A$. The couplings extracted through the kinetic term of the 2HDM Higgs Lagrangian are as follows:
\begin{align}
	HH^+W^-&:\frac{gs_{\beta-\alpha}}{2}(p^{\mu}_H-p^{\mu}_{H^+})\nonumber\\ 
	\h125H^+W^-&:\frac{gc_{\beta-\alpha}}{2}(p^{\mu}_{\h125}-p^{\mu}_{H^+})\nonumber\\ 
	AH^+W^-&:\frac{ig}{2}(p^{\mu}_{H^+}-p^{\mu}_A). 
	\label{HW}
\end{align}
In a recent work, we evaluated the signal observability of $H^+ \to W^+H^0$ at the alignment limit in a future lepton collider and concluded that such colliders are able to explore regions of parameter space not accessible by LHC \cite{Hashemi:2023}. 

The recent observations suggest that the experimental data are best described by a misaligned 2HDM if type I is the underlying theory. In such scenario, $H^+ \to W^+h_{125}$ can be a viable decay channel if $s_{\beta-\alpha}\neq1$. 

Such a decay channel competes with $H^+ \to t\bar{b}$ which is suppressed at high $t_\beta$ values in 2HDM type I due to couplings being proportional to $ct_\beta$.
 
The degenerate mass spectrum for this analysis is defined as $m_H=m_A=m_{H^+}$. This choice suppresses $H^+ \to W^+A$  and $H^+ \to W^+H$ in favor of $H^+ \to W^+h_{125}$.
\section{Constraints on the charged Higgs}
In this section, a review of different sources used to impose constraints on the charged Higgs mass and its couplings is presented. These can be divided into three main categories: theoretical constraints, indirect constraints from electroweak, flavor physics and $\h125$ at LHC, and direct search results from ATLAS and CMS which will all be discussed in detail. 
\subsection{Theoretical constraints}
The Higgs potential in Eq. \ref{potential} is constrained to theoretical requirements of positivity \cite{pos1,pos2,pos3,pos5}, unitarity and perturbativity \cite{uni1,uni2,uni3} and $\Delta\rho$ (to be consistent with electroweak precision measurements) \cite{drho1,drho2,drho3,drho4}. The chosen mass spectrum guarantees $\Delta\rho$ requirement. Other requirements of positivity, unitarity and perturbativity are checked and their excluded regions in the parameter space are mapped to the final plots.
\subsection{Electroweak precision measurement}
The observed deviation of the $W$ boson mass from SM by the CDF collaboration in 2022 \cite{Wmass} led to conclusions on the the mass splitting between the charged and neutral Higgs bosons \cite{DeltamH1,DeltamH2,DeltamH3,DeltamH4,DeltamH5, WmassTwoLoop}, upper limits of 1 TeV on their masses \cite{mWupperlimit,mWupperlimit2} and the muon $g-2$ anomaly \cite{mWg20,mWg21,mWg22,mWg23,mWg24}. However, results from LHCb \cite{LHCbW}, ATLAS \cite{ATLASW} and CMS \cite{CMSW} show reasonable agreement with electroweak global fit \cite{EWW}. 

\begin{figure}[hbt!]
	\centering
	\includegraphics[width=0.8\linewidth,height=0.7\linewidth]{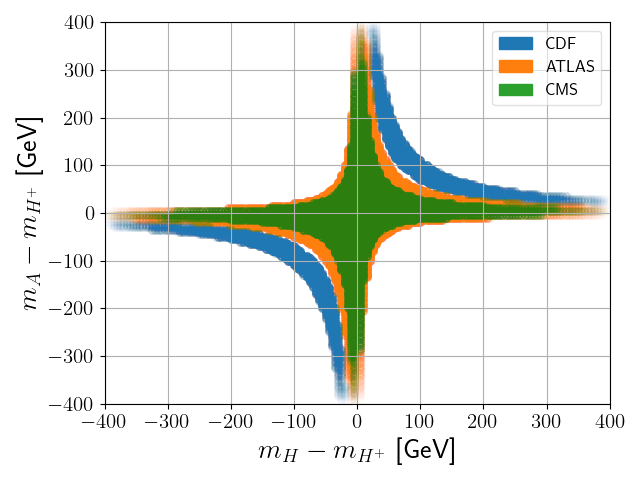}
	\caption{The mass splitting between the charged and neutral Higgs bosons consistent with the CDF, ATLAS and CMS $m_W$ measurements at 95$\%$ CL.}
	\label{dmw}
\end{figure}

In order to show the mass spectrum in terms of the neutral and charged Higgs boson mass splitting, we use the expression of $\Delta m_{W}$ in terms of $\Delta\rho$ which is a function of extra Higgs boson masses as in \cite{HBahlW}. The allowed regions consistent with the CDF measurement and recent reports from ATLAS and CMS are shown in Fig. \ref{dmw}. While CDF measurement urges for mass splitting between the neutral Higgs bosons and the charged Higgs, recent results from ATLAS and CMS allow the central region of equal masses favored by MSSM-like scenarios.     

\begin{figure}[hbt!]
	\centering
	\includegraphics[width=\linewidth,height=0.4\linewidth]{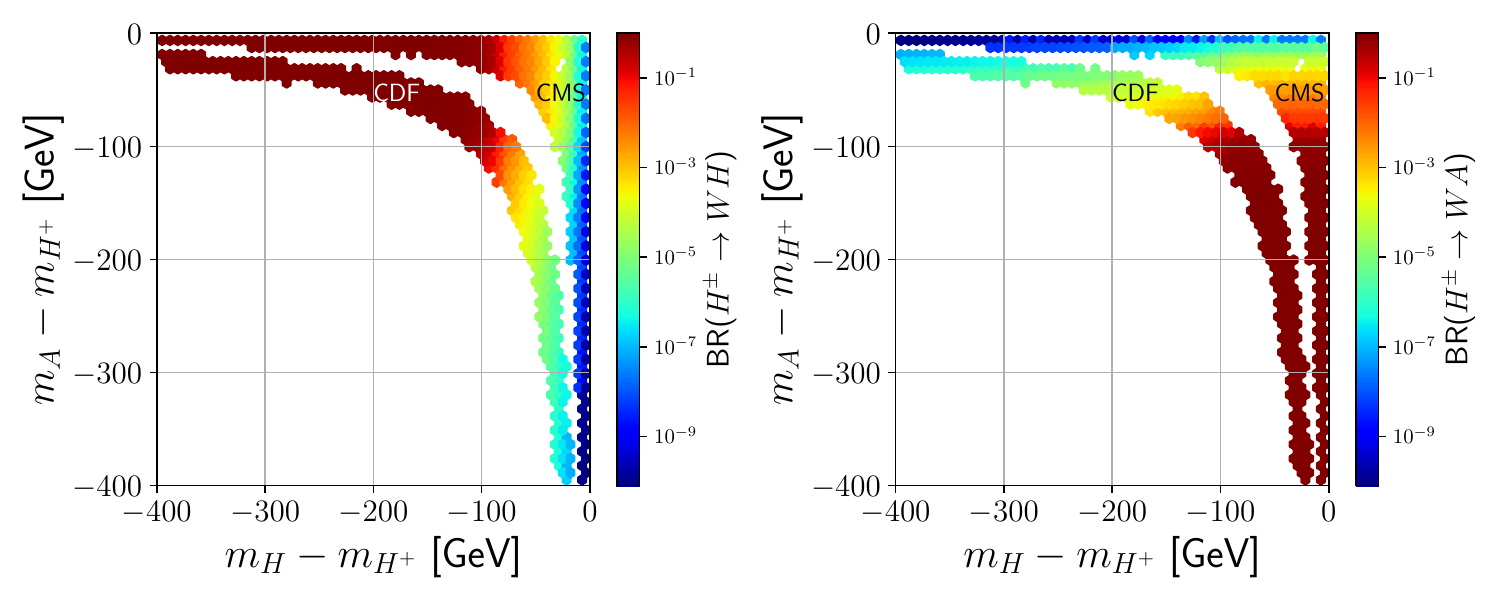}
	\caption{The branching ratio of charged Higgs decay to heavy neutral Higgs bosons H (left) and A(right) for the mass points consistent with the CDF and CMS $m_W$ measurements at 95$\%$ CL.}
	\label{dmw2}
\end{figure}

The mass splitting scenario could have resulted in off-shell decays of charged to heavy neutral Higgs bosons which would suppress the decay channel $\hpm \to W \h125$ under study in this work. Therefore the two channels $\hpm \to W^*H$ and $\hpm \to W^*A$ were studied in terms of their branching ratios with Higgs boson mass splittings allowed by CDF and the most recent result from CMS. Results are shown in Fig. \ref{dmw2} stating that a $\sim 10\%$ contribution from these channels forced by CDF, can be reduced to almost zero with mass degeneracy allowed by CMS.   
\subsection{Flavor physics constraints}
The precision measurement of flavor physics observables involves heavy meson decays and is sensitive to contributions from the charged Higgs bosons. Currently the $b\to s\gamma$ decay \cite{Misiak,Misiak2,Misiak3,Misiak4} provides strongest indirect limit on the charged Higgs mass and excludes masses below 600 GeV in 2HDM type II. The exclusions in 2HDM type I and X are very limited to regions below $\tb<2$, while in type Y similar region as in type II is excluded except for high $\tb$ \cite{FMahmoudi}.
\subsection{Direct searches at LHC (ATLAS and CMS)}
The experimental collaborations have divided their direct searches into two regions of light $(m_{H^+} \lesssim m_t)$ and heavy $(m_{H^+} \gtrsim m_t)$ charged Higgs bosons. The light charged Higgs boson can be produced through the top quark decay and has almost been fully excluded in 2HDM types II and X by CMS \cite{LightCH7TeVCMS1,LightCH7TeVCMS2,CHtaunu8TeVCMS,CMS:chnew6} and ATLAS \cite{LightCH7TeVATLAS1,LightCH7TeVATLAS2,CHtaunu8TeVATLAS,ATLAS:chnew1,ATLAS:2024taunu}. In 2HDM type Y, in the low mass charged Higgs scenario, the above searches by CMS and ATLAS show no sizable sensitivity because in this case $\hp \to c\bar{b}$ or $\hp \to c\bar{s}$ are the main channels in type Y and they escape from the searches for $\tau\nu$ or $tb$ in \cite{LightCH7TeVCMS1,LightCH7TeVCMS2,CHtaunu8TeVCMS,CMS:chnew6,LightCH7TeVATLAS1,LightCH7TeVATLAS2,CHtaunu8TeVATLAS,ATLAS:chnew1}. The 2HDM type I also allows a large part of parameter space for light and heavy charged Higgs bosons due to the $\ctb$ factor in the charged Higgs-fermion couplings and the exclusion is limited to low $\tb$ values. 

The heavy charged Higgs can undergo $H^+ \to t\bar{b}$ decay and masses above 200 GeV have been excluded for $\tan\beta$ values below 2.1 and above 34 by ATLAS \cite{ATLAS:chnew5,ATLAS:chnew2} and CMS \cite{CMS:chnew1,CMS:chnew4}. These results are based on MSSM with 2HDM type II as the Higgs basis.

Decays to lighter quarks have been analyzed through $H^+ \to c\bar{s}$ by CMS \cite{CMS:chnew3} and $H^+ \to c\bar{b}$ by CMS \cite{CMS:chnew7} and ATLAS \cite{ATLAS:chnew3,ATLAS:chnew4,ATLAS:2024cs}. These analyses focus on the light charged Higgs and have set upper limits on BR$(t \to \hp b)$ assuming BR$(\hp \to cs~ \tn{or}~cb)=1$.   

Apart from the charged Higgs decay to quarks, bosonic decay to $W$ boson has also been shown to be possibly sizable at $\tan\beta \sim 7$ \cite{CH2WhWH}. 

The LHC searches for such decays include $H^+ \to W^+H$ \cite{CMS:chnew2} and $H^+ \to W^+A$ \cite{CMScH2WA,CMScH2WA2} by CMS and $H^+ \to W^+Z$ by CMS \cite{CMScH2WZ} and ATLAS \cite{ATLAScH2WZ,ATLAScH2WZ2}. 

The CMS analysis of $H^+ \to W^+H$ reported in \cite{CMS:chnew2} provides upper limits on the charged Higgs production cross section times BR$(\hp \to WH)*$BR$(H \to \tau\tau)$ assuming a fixed value for the heavy neutral Higgs boson mass, i.e., $\mH=200$ GeV.  

The CMS analysis reported in \cite{CMScH2WA,CMScH2WA2} searches for $\hp \to WA$ with light charged Higgs signals produced in top quark decays leading to upper limits on BR$(t \to \hp b)*$BR$(\hp \to WA)$. 

Searches for $\hp \to W^+Z$ reported in \cite{CMScH2WZ,ATLAScH2WZ,ATLAScH2WZ2,ATLAS:2024GM} are based on Higgs-triplet scenarios such as Georgi-Machacek model \cite{GM}.
 
There have been search proposals for the charged Higgs boson decay to SM-like Higgs boson, i.e., $H^{\pm}\to W^{\pm}h_{125}$ at LHC/HL-LHC \cite{CH2WhLHC} and also $H^{\pm}W^{\mp}Z$ at HE-LHC \cite{CHWZ}. 

The light charged Higgs bosonic decays in 2HDM type I have been studied in \cite{CH2WhArhrib}. The associated production at LHC $pp\to H^{\pm}\phi$ with $\phi=h_{125}/H/A/W^{\mp}$ has been discussed in \cite{CH2WH}. There is also $H^{\pm}\to W^{\pm}h_{125}/A$ search proposal through associated or pair production of the charged Higgs at LHC \cite{CHLightNewModes}. 

There is a recent analysis performed by ATLAS collaboration focusing on the same decay channel as the topic of this work \cite{ATLAS:2024}. In their search for $\hp \to W\h125$, ATLAS collaboration set upper limit on the signal cross section times BR$(\hp \to W\h125)$*BR$(\h125 \to b\bar{b})$. 

\begin{figure}[hbt!]
	\centering
	\includegraphics[width=\linewidth,height=\linewidth]{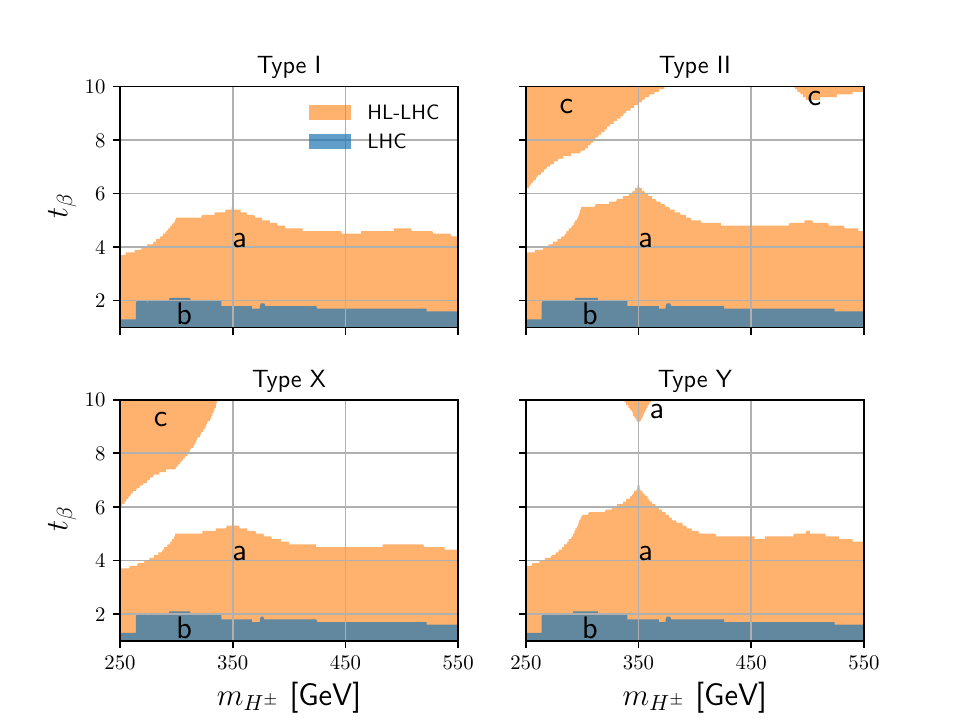}
	\caption{Charged Higgs excluded area at 95$\%$ CL at LHC in different 2HDM types. The HL-LHC exclusion potential is also shown for each type of the model. Labels refer to a: $H^{\pm}\to tb$ (CMS) \cite{CMS:chnew1}, b: $H^{\pm}\to tb$ (ATLAS) \cite{ATLAS:chnew2} and c: $H^{\pm}\to \tau\nu$ (ATLAS) \cite{ATLAS:chnew1}. The HL-LHC regions are obtained by extrapolating LHC analyses results to HL-LHC luminosity of 3000 $fb^{-1}$.}
	\label{CHalltypes}
\end{figure}

In order to summarize and provide a picture of the current status of experimental searches, we use \texttt{HiggsTools} and its recent database which includes results of 159 analyses of LEP and LHC (7, 8, 13 TeV). In Fig. \ref{CHalltypes} the LHC excluded regions in the parameter space relevant for the current analysis are shown for the four types of the model at the alignment limit. The HL-LHC expectation at integrated luminosity of 3000 $fb^{-1}$ is also shown based on the assumption that observed signal ratios follow statistical evolution with luminosity \cite{hllhc2}. 

Figure 3 shows that the most sensitive search channels at LHC and HL-LHC are $\hpm \to tb$ and $\hpm \to \tau\nu$. The region at low $\tb$ (labeled as $b$) is excluded at LHC with $\hpm \to tb$ search by ATLAS \cite{ATLAS:chnew2}. The HL-LHC is expected to be able to exclude regions labeled as $a$ and $c$ through $\hpm \to tb$ \cite{CMS:chnew1} and $\hpm \to \tau\nu$ \cite{ATLAS:chnew1} respectively. Therefore the most sensitive charged Higgs decay channels at LHC are fermionic decays at the alignment limit. There is a very recent result from ATLAS reported in \cite{ATLAS:2024CH} where $\hpm \to \tau\nu$ has been analyzed in type II and extends excluded region of type II to lower $\tb$ values which were previously reported in \cite{ATLAS:chnew1}. This decay channel is more relevant to type II at high $\tb$. In type I, the dominant channel is $\hpm \to tb$ at the alignment limit. Moreover, the aim of this work is to analyze the mis-aligned 2HDM where the charged Higgs decay to gauge bosons are enhanced and fermionic decays are suppressed.

\begin{figure}[hbt!]
	\centering
	\includegraphics[width=0.8\linewidth,height=0.7\linewidth]{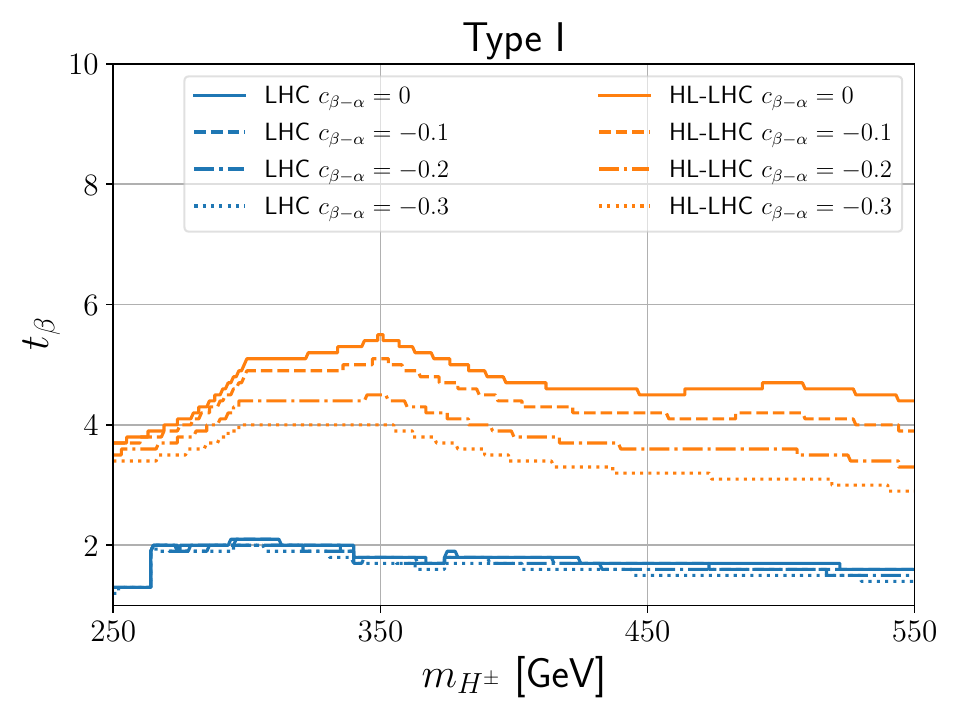}
	\caption{The effect of deviation from the alignment limit on the charged Higgs excluded area at LHC in 2HDM type I. The HL-LHC expectation is also shown. The largest areas in both scenarios of LHC and HL-LHC belong to the case of $\cba=0$ (alignment limit).}
	\label{CHtype1}
\end{figure}

\begin{figure}[hbt!]
	\centering
	\includegraphics[width=\linewidth,height=\linewidth]{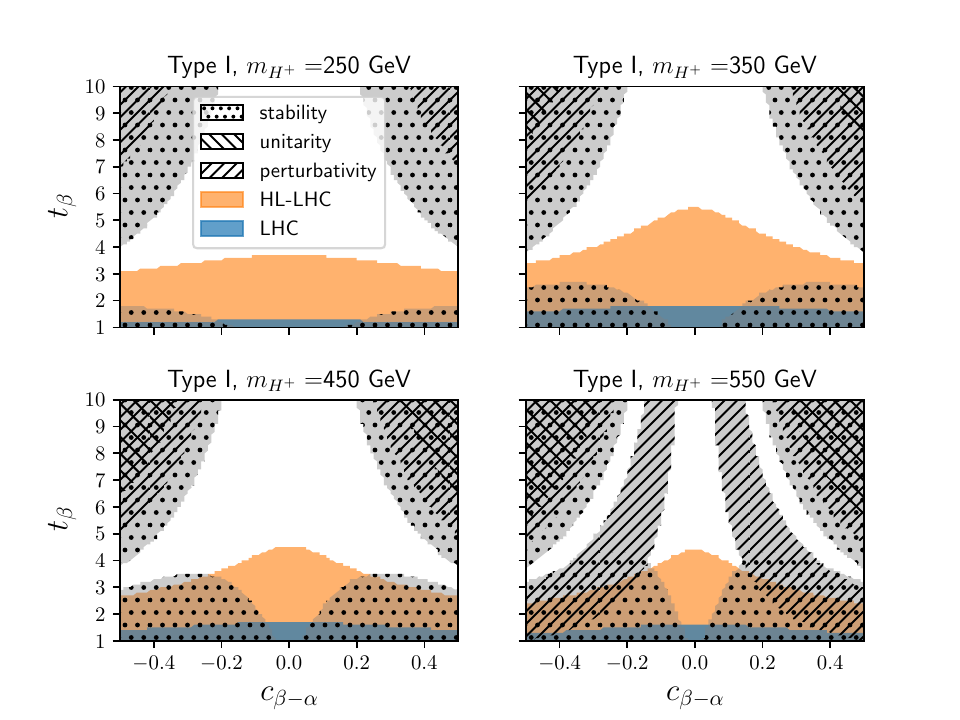}
	\caption{The theoretical exclusion of the parameter space in 2HDM type I for four charged Higgs masses 250 to 550 GeV including individual results from stability, unitarity and perturbativity. The LHC excluded area at 95$\%$ CL and HL-LHC expectation are also shown.}
	\label{sigma_bare}
\end{figure}

The effect of deviation from the alignment limit is illustrated in Fig. \ref{CHtype1} in terms of the coupling factor $\cba$. The farther from the alignment limit, the softer the exclusion. In a different view, we show the excluded regions in $\cba,\tb$ space including theoretically excluded regions by stability, unitarity and perturbativity requirements in Fig. \ref{sigma_bare}. The HL-LHC expectation at 3000 $fb^{-1}$ is also shown with the same orange color as previous figures. The chosen charged Higgs masses are benchmark points used for the event analysis.
  
The conclusion is that despite the extensive search for the charged Higgs decays, a large region of the parameter space is still unexplored in misaligned 2HDM scenario. The aim of this work is to provide a parameter space scan in terms of $c_{\beta-\alpha}$ and $\tb$ in 2HDM type I and discuss about heavy charged Higgs observability in such scenario through $\hp \to W\h125$ searches.
\section{Analysis of $H^+ \to W^+h_{125}$ in misaligned 2HDM type I}
Before describing the analysis, a list of required software is provided followed by a motivation to choose misaligned 2HDM type I as the basis for the analysis. 
\subsection{Software setup}
The following public software are used to do the analysis: 
\begin{itemize}
	\item \texttt{2HDMC-1.8.0} for decay rates and branching ratios of neutral and charged Higgs bosons and theoretical constraints \cite{2hdmc1,2hdmc2,2hdmc3}, 
	\item \texttt{HiggsTools-1} for experimental results from LEP to Tevatron and LHC \cite{higgstools}, 
	\item \texttt{FeynRules-2} for 2HDM type I model file generation \cite{feynrules,ufo},
	\item \texttt{MadGraph5\_aMC@NLO-3.6.2} \cite{mg51,mg52} for LHC and \texttt{WHIZARD-3.1.2} \cite{whizard1,whizard2} for lepton collider event cross section calculation and hard scattering generation,
	\item \texttt{circe2} for lepton colliders beam spectra \cite{circe2}, 
	\item \texttt{PYTHIA-8.3.09} for multi-particle interactions and final state radiation and showering \cite{pythia}, 
	\item \texttt{DELPHES-3.5.0} for detector simulation \cite{delphes1,delphes2,delphes3},
	\item \texttt{delphes\_card\_CMS} and \texttt{CLICdet\_Stage2} for parameter definition of the physical object reconstruction algorithms \cite{tdr1,tdr2,clicdp,overlay,clicdet},
	\item \texttt{python3} libraries \texttt{numpy} \cite{numpy} and \texttt{matplotlib} for final result visualization \cite{matplotlib}.
\end{itemize}
\subsection{Motivation}
\begin{figure}[hbt!]
	\centering
	\includegraphics[width=\linewidth,height=\linewidth]{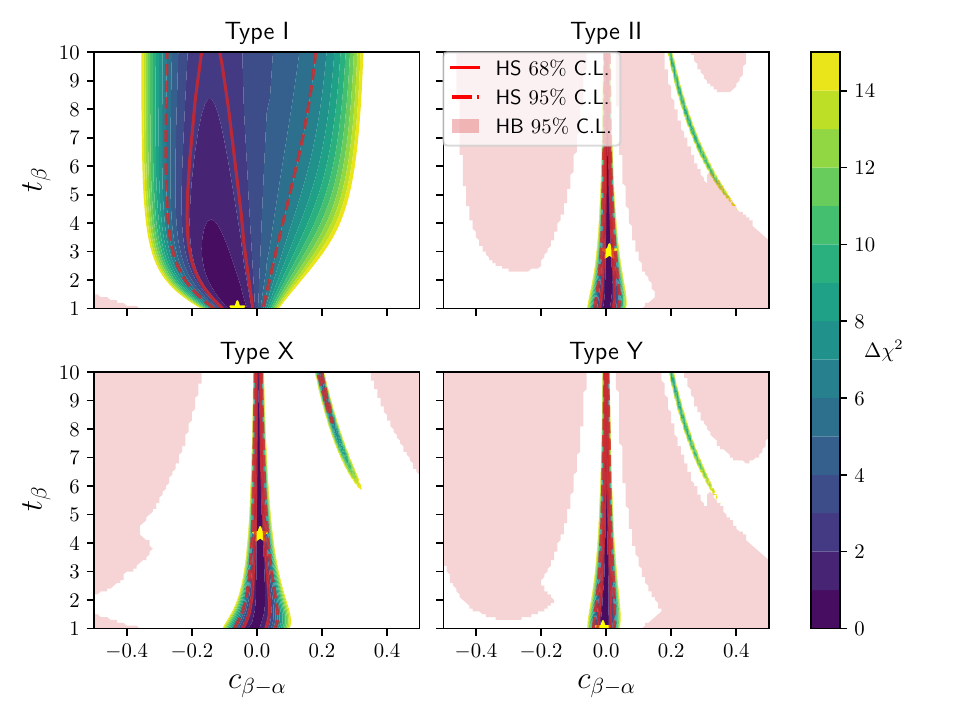}
	\caption{$\Delta\chi^2$ of the 2HDM types with respect to the best fit point with minimum $\chi^2$ shown with star. The red solid and dashed lines indicate the area of 68$\%$ and 95$\%$ CL around the best fit point. The light red area shows $h_{125}$ exclusion at 95$\%$ CL by the current LHC data.  }
	\label{chi2}
\end{figure}

Using $h_{125}$ precision measurement data from LHC, the model deviation from its best fit to data is expressed in terms of $\Delta\chi^2$ variable which we obtain using \texttt{HiggsTools} and its \texttt{HiggsSignals} (HS) component. Results are shown in Fig. \ref{chi2} for the four types of 2HDM. The light red area has been obtained using \texttt{HiggsBounds} (HB) and shows the current LHC excluded area at 95$\%$ CL. 

The alignment limit is not excluded and is preferred in model types II, X and Y. However, in type I, the best fit and the area of low $\Delta\chi^2$ appear in negative $c_{\beta-\alpha}$ values as seen in Tab. \ref{c2}. 
\begin{table}[h]
	\centering
	\begin{tabular}{|c|c|c|c|c|}
		\hline
		2HDM type & I & II & X & Y\\
		\hline
		$\cba,~\tb$ & -0.06, 1 & 0.01, 3 &  0.01, 4 & -0.01, 1  \\
		\hline
		$\chi^2_{\tn{min}}/n.d.f$ & 149.0/159 & 152.0/159 & 151.9/159 & 152.3/159  \\
		\hline
	\end{tabular}
	\caption{Coordinates of the best fit point with minimum $\chi^2$ in parameter space of $\cba$ vs. $\tb$ for different 2HDM types. The corresponding SM value is $\chi^2/n.d.f=152.54/159$. The total number of 159 analyses from LEP and LHC 8 and 13 TeV runs are included.}
	\label{c2}
\end{table}

This observation indicates that if the observed Higgs boson at LHC belongs to a 2HDM type I, non-SM couplings are preferred and make a better agreement with LHC observation. Moreover, the minimum $\chi^2$ obtained in 2HDM type I is the absolute minimum among all 2HDM types and also SM. Therefore a misaligned 2HDM type I agrees best with the current LHC Higgs boson precision measurement among other scenarios. 

In what follows, we present two separate analyses at (HL-)LHC and lepton collider (LC). Although the signal to background cross section ratio is small at LHC, we perform an event generation and simulation for a more detailed and quantitative result.   
\subsection{Choice of the collider}
The mass of a charged Higgs boson should be above 205 GeV to decay through $\hpm \to \wpm h_{125}$. Such a heavy charged Higgs can be produced in proton-proton collisions at LHC and its future high luminosity and high energy upgrades (HL-LHC and HE-LHC) \cite{hllhc1,hllhc2} and later on at Future Circular Collider (FCC-hh) with $\sqrt{s}=100~\TeV$ \cite{FCC-hh}. 

Pair production at $e^-e^+$ colliders through $e^-e^+ \to H^+H^-$ needs center of mass energies in the range 400 to 1400 GeV for a charged Higgs up to 700 GeV. This requirement excludes FCC-ee with $\sqrt{s}=350~\GeV$ \cite{FCC-ee1,FCC-ee2}, the International Linear Collider (ILC) with $\sqrt{s}=500~\GeV$ \cite{ILC,ILCEnergy,ILCEnergy2} and CEPC with $\sqrt{s}=240 ~\GeV$ \cite{CEPC1,CEPC2}. 

While LHC is proceeding with charged Higgs search programs, lepton colliders are alternative scenarios due to their cleaner collision environment which results in better signals on top of the background processes. Studies of such colliders and their potential for extra Higgs boson discovery include ILC \cite{ilchiggs} and CEPC \cite{cepchiggs}. These results are limited by the center of mass energy of the colliders. A reasonable choice for high energy collisions is Compact Linear Collider (CLIC) which is expected to operate in second stage at $\sqrt{s}=1400$ GeV \cite{clichiggs1,clichiggs2}. The CLIC experiment will provide the first lepton collisions at such energy and provides an opportunity to study the charged Higgs boson pair production with $m_{\hp}>200$ GeV through $e^-e^+$ collisions. 

Since the charged Higgs pair production is also possible at LHC, we first perform an analysis of $pp \to H^+H^-$ to evaluate its performance compared to $pp \to tbH^\pm$ at LHC and the corresponding pair production at LC.          
 
\subsection{Decay branching ratios of $h_{125}$ and $\hpm$}
\begin{figure}[hbt!]
	\centering
	\includegraphics[width=\linewidth,height=\linewidth]{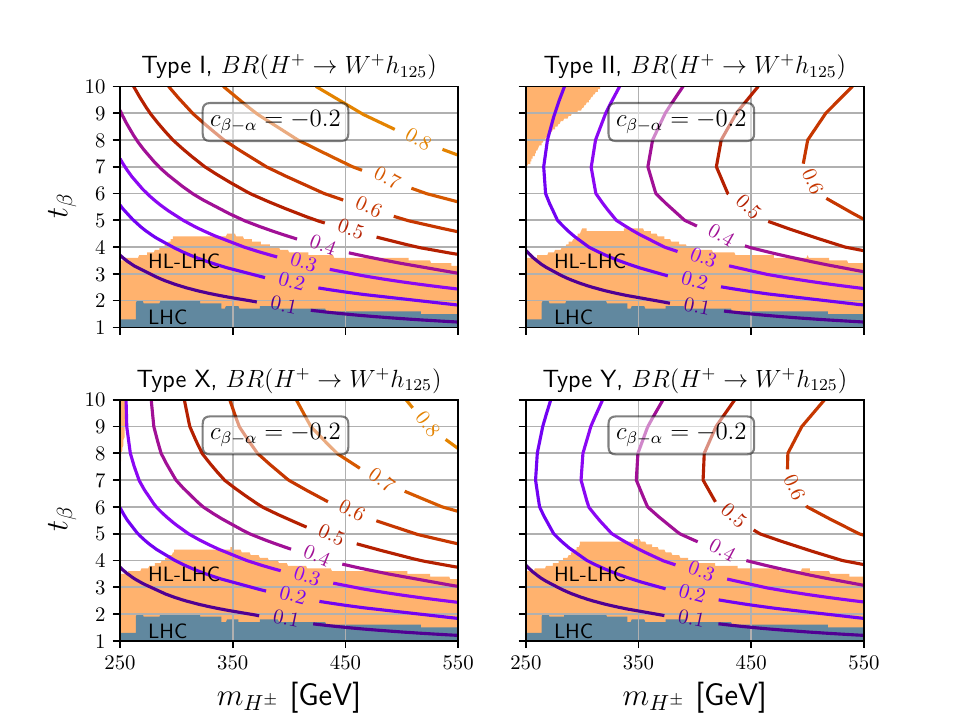}
	\caption{Branching ratio of charged Higgs decay to $W\h125$ in different 2HDM types for a fixed value of $\cba=-0.2$. The LHC excluded region at 95$\%$ CL and HL-LHC expectation at 3000 $fb^{-1}$ are also shown.}
	\label{brwh}
\end{figure}

\begin{figure}[hbt!]
	\centering
	\includegraphics[width=\linewidth,height=\linewidth]{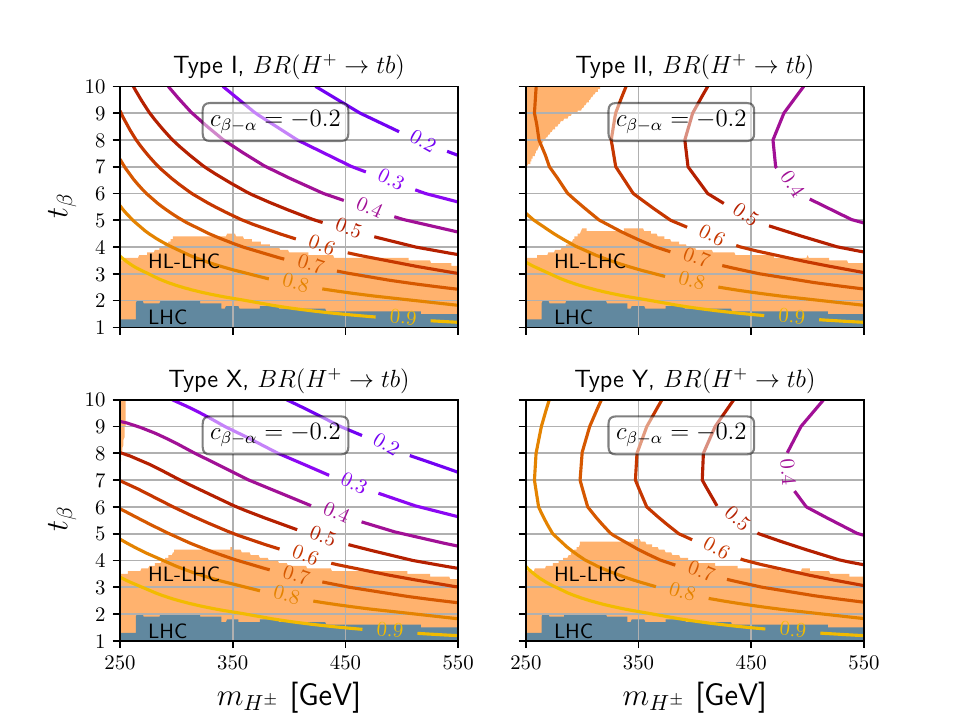}
	\caption{Branching ratio of charged Higgs decay to $tb$ in different 2HDM types for a fixed value of $\cba=-0.2$.}
	\label{brtb}
\end{figure}

\begin{figure}[hbt!]
	\centering
	\includegraphics[width=\linewidth,height=\linewidth]{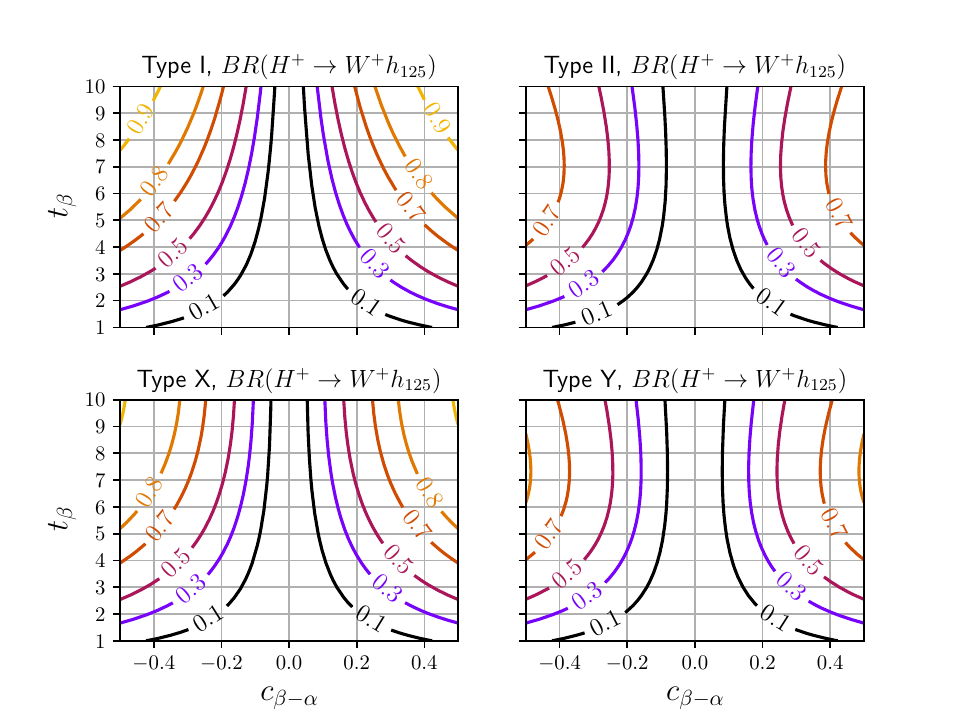}
	\caption{Branching ratio of charged Higgs decay to $W\h125$ in different 2HDM types as a function of $\tb,~\cba$ for a fixed value of $\mhpm=350$ GeV. }
	\label{brch}
\end{figure}

The charged Higgs boson decay $H^+\to Wh_{125}$ occurs at non-zero $\cba$ and competes with $H^+ \to tb$. We compare the two channels in Figs. \ref{brwh} and \ref{brtb} as a function of the charged Higgs mass and $\tb$ for a fixed value of $\cba=-0.2$. The negative value of $\cba$ is chosen due to type I preference. The excluded region of LHC at 95$\%$ CL and HL-LHC expectation are also shown in both Figs. \ref{brwh} and \ref{brtb}. As is seen, $H^+ \to W\h125$ prefers high $\tb$ and $\mhpm$ region, while $H^+ \to tb$ prefers light charged Higgs or low $\tb$ region. Therefore one advantage of searching for $H^+\to W\h125$ is the enhancement of this decay with increasing charged Higgs mass at a fixed $\tb$. The charged Higgs boson branching ratio of decay through $H^+\to Wh_{125}$ can be shown in $\tb,~\cba$ but for a fixed mass. An example with $\mhpm=350$ GeV is shown in Fig. \ref{brch} which shows larger branching ratios with increasing $\cba$ for any fixed value of $\tb$.

\begin{figure}[hbt!]
	\centering
	\includegraphics[width=\linewidth,height=\linewidth]{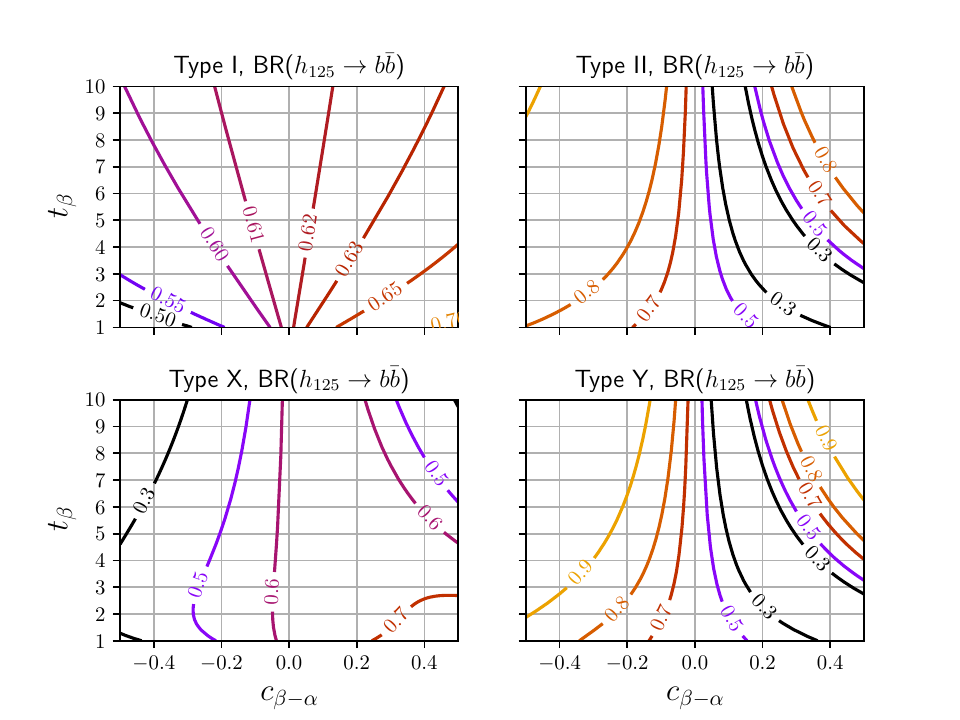}
	\caption{Branching ratio of light neutral Higgs decay to $bb$ in different 2HDM types as a function of $\tb,~\cba$.}
	\label{brh}
\end{figure}

The $h_{125}$ couplings to down-type quarks relevant for $h_{125} \to bb$ analysis are listed in Tab. \ref{couplings} and are in general functions of $\alpha$ and $\beta$. The decay branching ratio has been plotted as a function of $\tb$ and $\cba$ in Fig. \ref{brh}.
\PartTitle{Part I}
\subsection{Search strategy and methodology at (HL-)LHC}
\begin{figure}[hbt!]
	\centering
	\includegraphics[width=0.6\linewidth,height=0.55\linewidth]{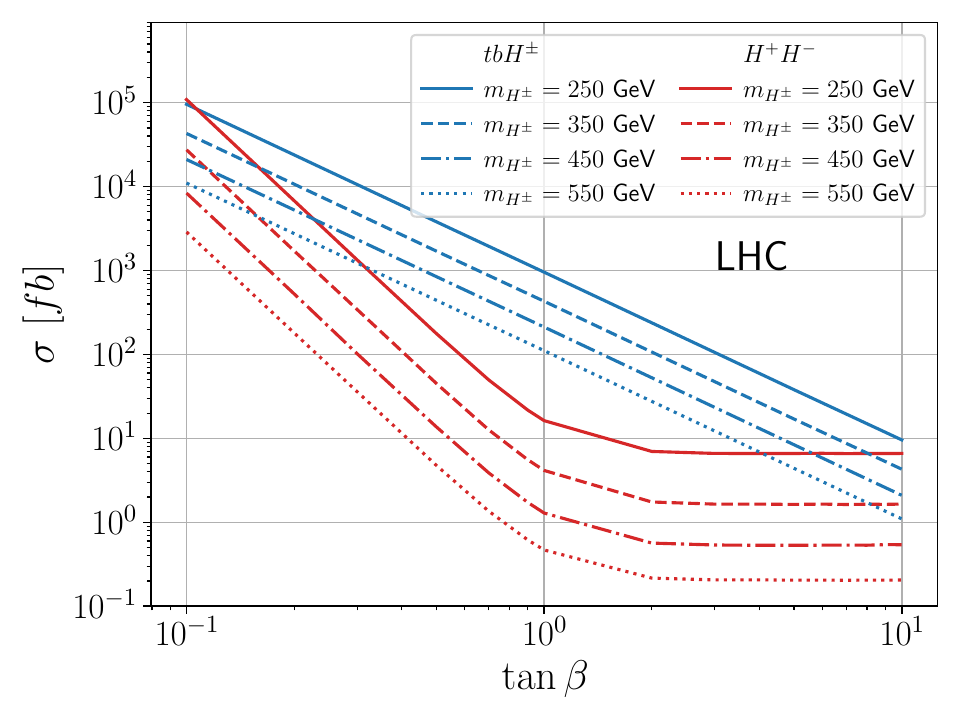}
	\caption{Charged Higgs production cross section through $pp \to tbH^{\pm}$ and $pp \to H^+H^-$ at LHC with $\sqrt{s}=13$ TeV as a function of $\tb$ for different charged Higgs masses.}
	\label{compare}
\end{figure}

The signal under consideration at LHC is $q\bar{q} \to \gamma/Z/h_{125}/H \to H^+H^-$  in $s-$channel as well as $b\bar{b} \to H^+H^-$ in $t-$channel and loop initiated gluon-gluon fusion. The latter is estimated to have a $10\%$ contribution to the total cross section \cite{gg2HpHm1,gg2HpHm2}. The $s-$channel vertices are \cite{arh}:
\begin{align}
	\nonumber&g_{\h125H^+H^-}:\\ 
	\nonumber&\frac{1}{v}\left[ \left(2\mhpm^2-\mh^2\right)\sba + \left( \mh^2-2\frac{\m12^2}{s_{2\beta}}\right)\frac{c_{\beta+\alpha}}{\sb\cb}\right]\\ 
	\nonumber&g_{HH^+H^-}:\\
	\nonumber&\frac{1}{v}\left[ \left(2\mhpm^2-\mH^2\right)\cba + \left( \mH^2-2\frac{\m12^2}{s_{2\beta}}\right)\frac{s_{\beta+\alpha}}{\sb\cb}\right]\\ 
	\nonumber&g_{\gamma H^+H^-}:ie\\ 
	\nonumber&g_{ZH^+H^-}:i\frac{g(\cos^2\theta_W-\sin^2\theta_W)}{2\cos\theta_W}\\ 
	\label{cch}
\end{align}
The first two are functions of $\mh,\mH,\mhpm,\alpha,\beta$ and $\m12^2$ and are negligible due to $1/v$ factor. The $t-$channel diagrams are suppressed at high $\tb$ in type I. Therefore the signal cross section starts from largest values at low $\tb$ (e.g. 16$fb$ at $\tb=1$ for $\mhpm=250$ GeV) and decreases and reaches constant values when the contribution from diagrams containing Higgs-fermion couplings (the $t-$channel and loop diagrams) is suppressed. Figure \ref{compare} makes a comparison between the two production processes and shows that for the $\tb$ range under study, the LHC search channel ($tbH^{\pm}$) has a higher cross section. 

The main SM background processes are $t\bar{t}$ \cite{ttNLO1,ttNLO2}, $t\bar{t}b\bar{b}$ \cite{ttbb1}, single top production ($s$-channel $t\bar{b}$, $t$-channel $tq$ and $tW$)\cite{st} and vector boson pair production $VV$($W^+W^-,~W^{\pm}Z,~ZZ$)\cite{CMSWW,ATLASWW,CMSWZ,ATLASWZ,CMSZZ,ATLASZZ} whose cross sections together with the signal with different charged Higgs masses are listed in Tab. \ref{LHCxsecs}. 

For the signal cross section calculation we generate a UFO model file \cite{ufo} with \texttt{FeynRules} \cite{feynrules} for 2HDM type I and use it in \texttt{MadGraph5\_aMC@NLO} with built-in \texttt{NNPDF-2.3}\cite{nnpdf} as the parton distribution function. The single gauge boson production $Z/\gamma$ and $W+$jets are not considered as expected to be suppressed with jet multiplicity and $b$-tagging requirements. Other processes of associated production of top quark and $W/Z/\h125$ have very small cross sections to contribute in the signal region.
\begin{table}[h]
	\centering
	\begin{tabular}{|c|c|c|c|c|}
		\hline
		\multicolumn{5}{|c|}{LHC, Signal}\\
		\hline
		$\mhpm$ [GeV] & 250 & 350 & 450 & 550 \\
		\hline
		$\sigma$ [fb] & 6.6 & 1.6 & 0.5 & 0.2 \\
		\hline
	\end{tabular}
	\vspace{0.1cm}
	\begin{tabular}{|c|c|c|c|c|}
		\hline
		\multicolumn{5}{|c|}{LHC, Background}\\
		\hline
		Process & $t\bar{t}$ & $t\bar{t}b\bar{b}$ & single top & $VV$ \\
		\hline
		$\sigma$ [pb] & 832 & 4.7 & 299 & 182 \\
		\hline
	\end{tabular}
	\caption{Cross sections of the signal and background samples at LHC at $\sqrt{s}=13$ TeV.}
	\label{LHCxsecs}
\end{table}

\begin{figure}[hbt!]
	\centering
	\includegraphics[width=\linewidth,height=\linewidth]{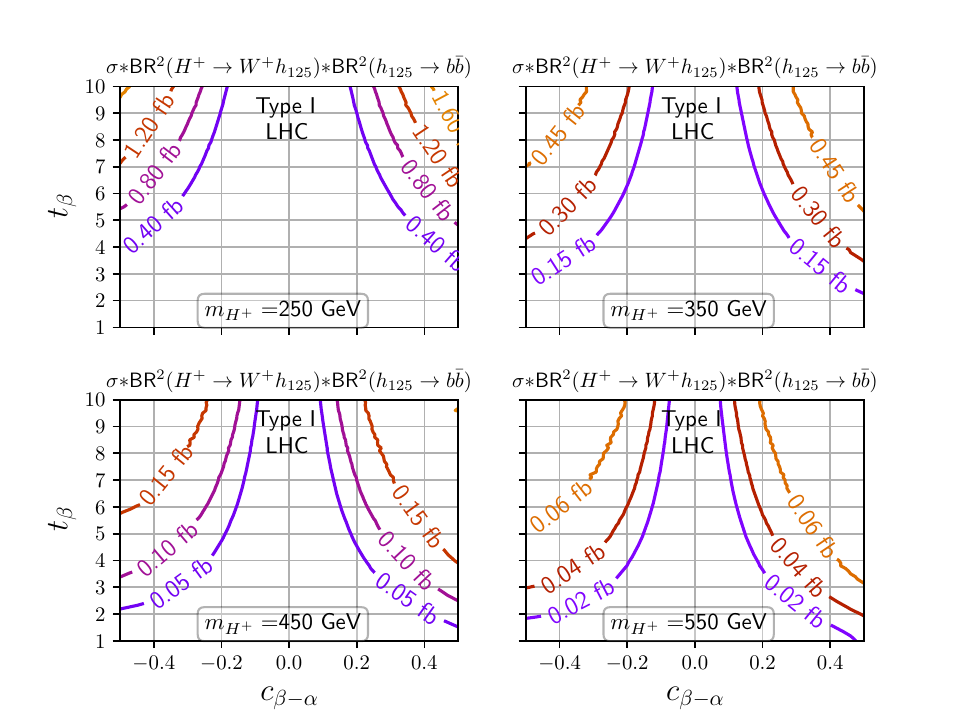}
	\caption{The product of the signal cross section at LHC times branching ratios of the charged and neutral Higgs decays in 2HDM type I as a function of $\tb,~\cba$. Results are shown for four masses $\mhpm=250,~350,~450,~550$ GeV.}
	\label{LHCbrchh}
\end{figure}

The product of the signal cross section at LHC times branching ratios of the charged and neutral Higgs decays is shown in Fig \ref{LHCbrchh} for four chosen charged Higgs masses in 2HDM type I. The initial cross sections times BRs are thus mostly below 1$fb$ in the parameter space under study.
When the charged Higgs pair is produced and $\hpm \to \wpm \h125$ occurs, the signal final state can be analyzed in fully hadronic final state where both $W$ bosons decay to light jet pairs, or semi-leptonic final state where one of the $W$ bosons decays hadronically and the other leptonically. The latter choice has the advantage of using lepton trigger for online event selection and reduced QCD multi-jet background. As for the light Higgs we consider $\h125 \to b\bar{b}$. This type of final state consists of 4 $b$-jets from two $\h125$ bosons, 2 light jets from one of the $W$ bosons and a single lepton (electron or muon, we do not take $\tau$ due to its hadronic decay), and missing transverse energy from neutrino.
  
The signal and background events are generated using \texttt{MadGraph5\_aMC@NLO} at 13 TeV LHC. The output event files in \texttt{LHEF} format \cite{lhef} are used by \texttt{PYTHIA} for final state showering and multi-particle interactions with detector simulation using \texttt{DELPHES} based on detector card \texttt{delphes\_card\_CMS}. 

The jet reconstruction algorithm anti-$k_T$ is used with kinematic acceptance 
\begin{equation}
	p_T^{\tn{jet}}>10 ~\GeV, ~~|\eta^{\tn{jet}}|<3
\end{equation} 
applied on the jet transverse momentum and pseudo-rapidity. This algorithm has been shown to be a natural, fast, infrared and collinear safe replacement for the previous iterative cone algorithms for hadron-hadron collision environment \cite{antikt}.

The jet separation is also applied by requiring all jet pairs to be separated by $\Delta R\geq0.4$ where $\Delta R = \sqrt{(\Delta \eta)^2+(\Delta \phi)^2}$. Here, $\eta$ and $\phi$ are pseudo-rapidity and azimuthal angle. The efficiency of this requirement is very close to unity for all signal and background samples.
 
The $b$-tagging efficiency is taken as a function of the jet transverse momentum and is ~70$\%$ for a $b$-jet with $p_T=100$ GeV. The mis-tagging rate for $c$-jets is also a function of the jet transverse momentum and is ~20$\%$ for a $c$-jet with $p_T=100$ GeV. The $p_T$ dependent light jet miss-tagging rate as a $b$-jet reaches ~1$\%$ for a light jet with $p_T=100$ GeV. If the $b$-tagging algorithm fails for a jet, the jet is assumed to be a light jet. 

\begin{figure}[hbt!]
	\centering
	\includegraphics[width=0.6\linewidth,height=.6\linewidth]{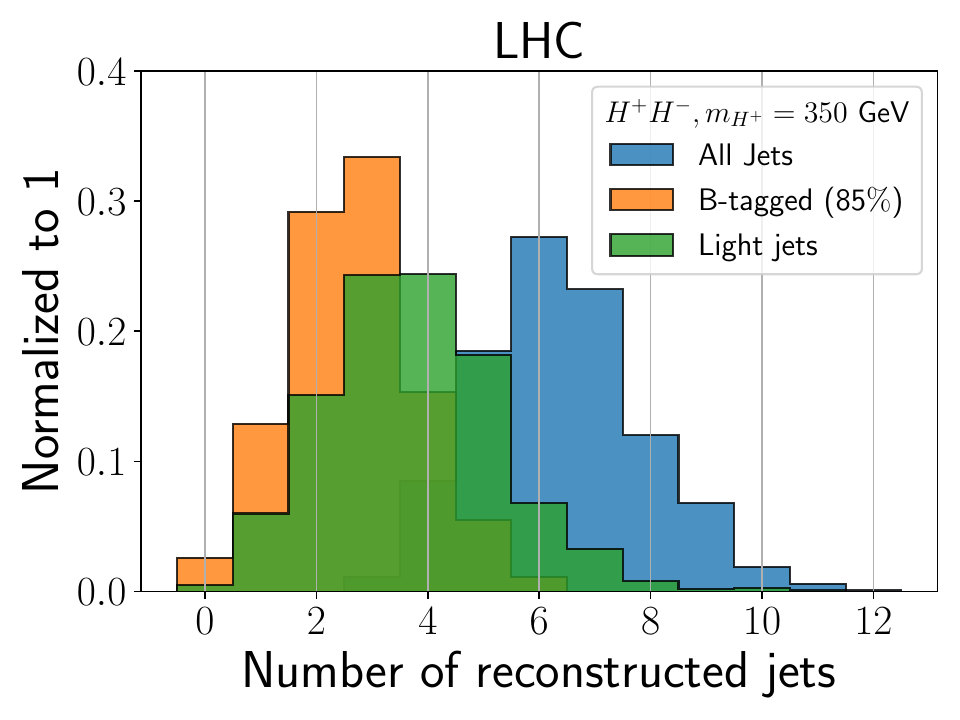}
	\caption{Number of $b$-jets, light jets and total number of jets in signal events with $\mhpm=350$ GeV.}
	\label{LHCSJetMul}
\end{figure}

\begin{figure}[hbt!]
	\centering
	\begin{subfigure}{.23\textwidth}
		\includegraphics[width=\linewidth,height=\linewidth]{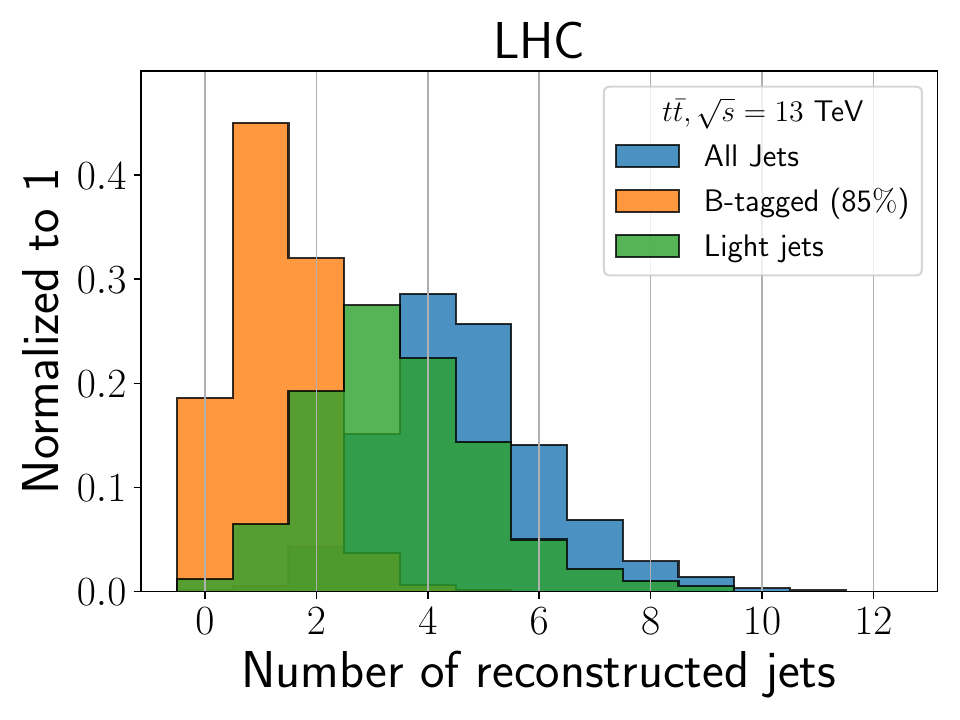}
		\caption{}
		\label{LHCttJetMul}
	\end{subfigure}%
	\begin{subfigure}{.23\textwidth}
		\includegraphics[width=\linewidth,height=\linewidth]{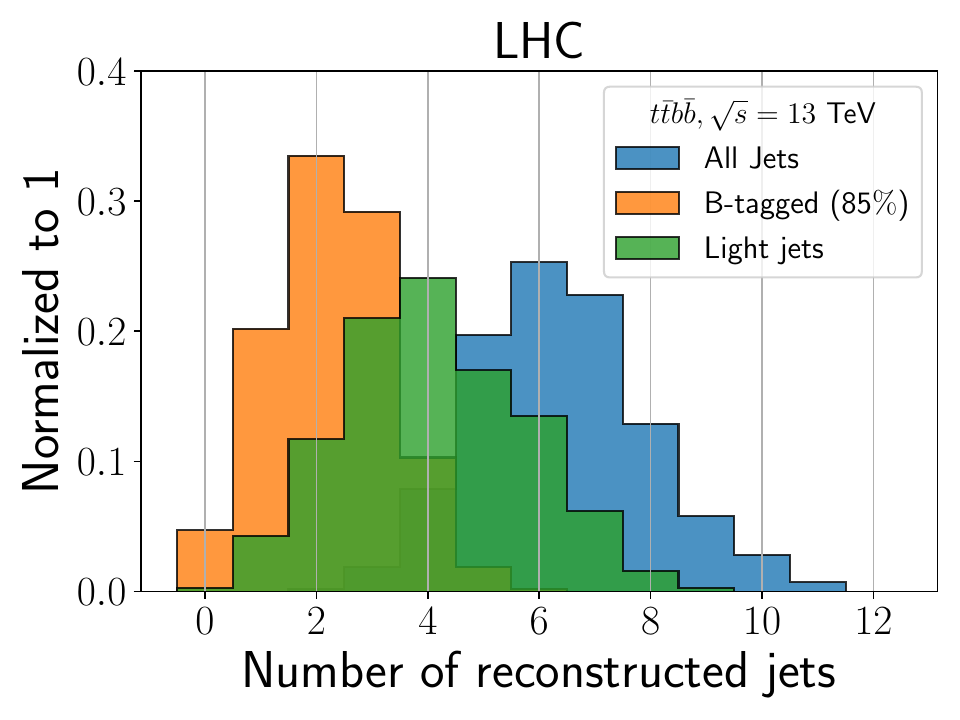}
		\caption{}
		\label{LHCttbbJetMul}
	\end{subfigure}
	\caption{Number of $b$-jets, light jets and total number of jets in $t\bar{t}$ (a) and $t\bar{t}b\bar{b}$ events (b).}
\end{figure}

The analysis starts with requiring at least two light jets and four $b$-jets in each event. The jet multiplicity distributions are shown in Fig. \ref{LHCSJetMul} for the signal ($\mhpm=350$ GeV) and Figs. \ref{LHCttJetMul} and \ref{LHCttbbJetMul} for $t\bar{t}$ and $t\bar{t}b\bar{b}$. 
The total number of jets is as expected in each plot, however, due to the low $b$-tagging efficiency for 4 $b$-jet final state, number of $b$-jets are less than four leading to enhancement of identified light jets. These distributions can be optimized with tuning the jet kinematic cuts and $b$-tagging parameters.

\begin{figure}[hbt!]
\centering
\includegraphics[width=0.95\linewidth,height=0.45\linewidth]{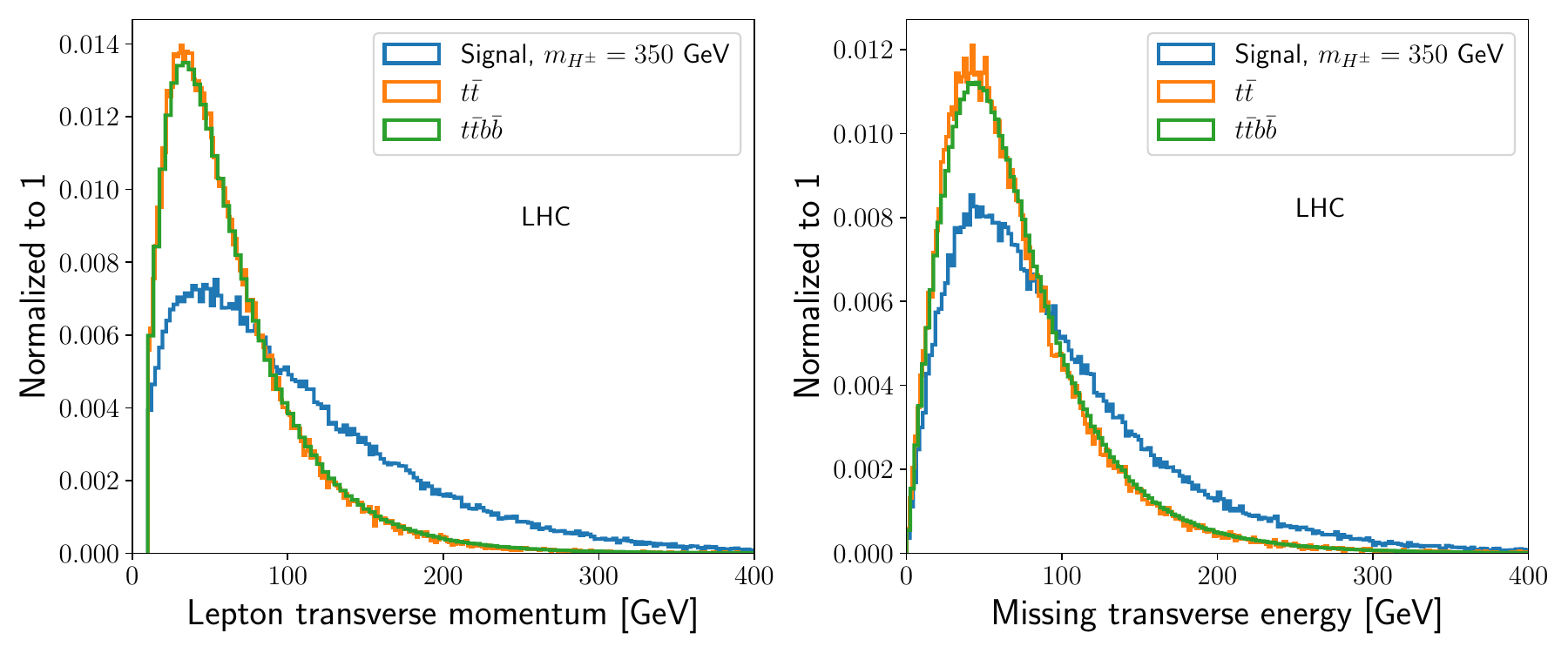}
\caption{The lepton (left) and neutrino (right) transverse momentum distributions in signal and the main SM background events.}
\label{letmet}
\end{figure}

In the next step, the lepton (electron or muon) and missing $E_{T}$ are reconstructed and accepted if the following kinematic cuts are satisfied:
\begin{align}
	p_T^{\text{lepton}} &> 10~\text{GeV}, \quad |\eta^{\text{lepton}}| < 3 \label{eq:let} \\
	\text{Missing } E_T &> 30~\text{GeV} \label{eq:met}.
\end{align}
There should be at least one lepton passing Eq. \ref{eq:let}. The kinematic distributions of leptons and missing $E_T$ in signal and the main background samples are shown in Fig. \ref{letmet}. Both distributions show that the signal is separated at about 80 GeV. The kinematic cuts of Eqs. \ref{eq:let} and \ref{eq:met} raised to this point but did not result in a higher signal significance. 

The $z$-component of the neutrino momentum ($p_Z^{\text{neutrino}}$) is reconstructed by requiring the invariant mass of the lepton+neutrino system to be equal to the $W$ boson mass. This requirement results in a second-degree polynomial equation. If no solution is found, $p_Z^{\text{neutrino}}$ is set to zero, otherwise the minimum solution is taken similar to \cite{ATLAS:2024}. 
 
\begin{figure}[hbt!]
	\centering
	\includegraphics[width=0.95\linewidth,height=0.45\linewidth]{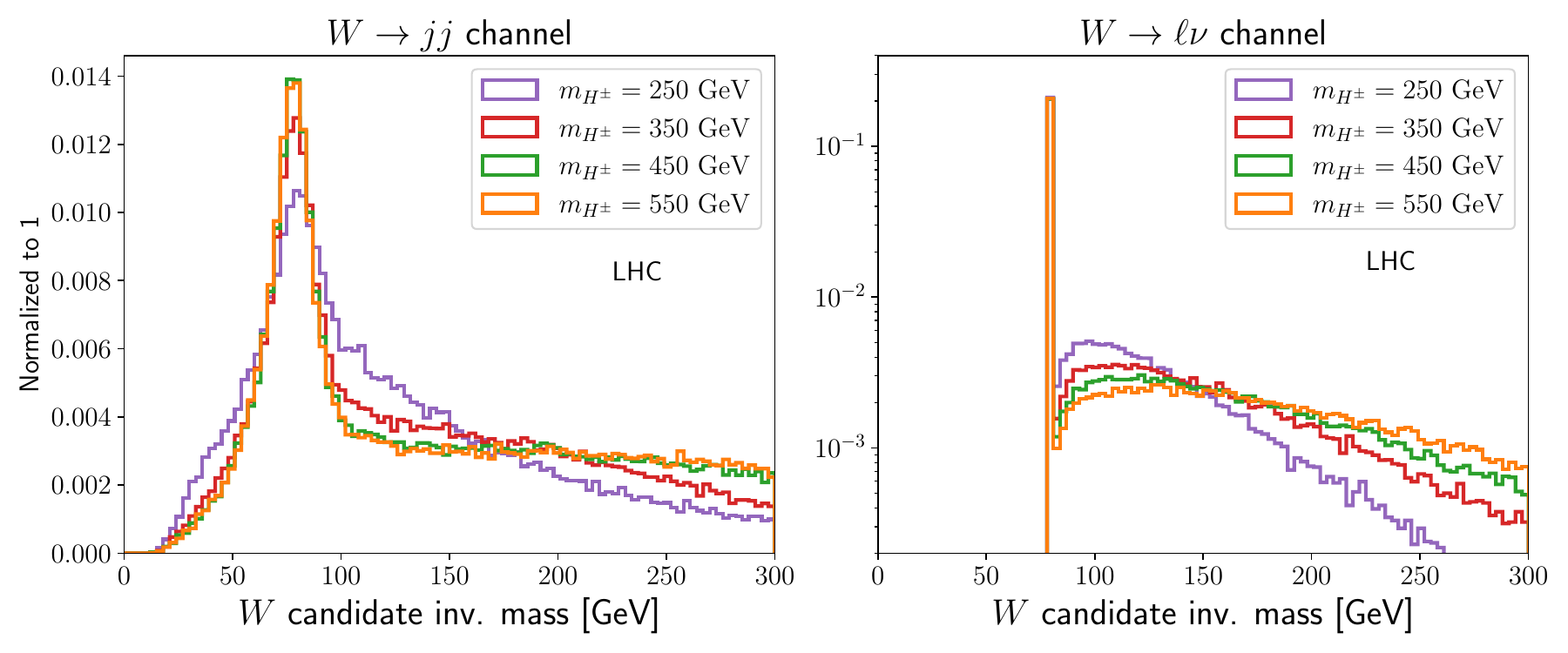}
	\caption{Invariant mass distributions of the reconstructed $W$ bosons in hadronic (left) and leptonic (right) decays in signal events.}
	\label{LHCSWs}
\end{figure}

The pair of light jets are then used to reconstruct hadronically decaying $W$ boson. Figure \ref{LHCSWs} shows reconstructed $W$ boson invariant mass distributions in hadronic and leptonic decay modes. The mass window for the reconstructed $W$ boson in the hadronic decay is $m_W\pm 20$ GeV and in the leptonic decay is $m_W\pm 5$ GeV which is trying to keep the tail of reconstructed $m_W$ distribution belonging to the cases where no solution was found for $p_Z^{\text{neutrino}}$.

\begin{figure}[hbt!]
	\centering
	\includegraphics[width=0.6\linewidth,height=0.55\linewidth]{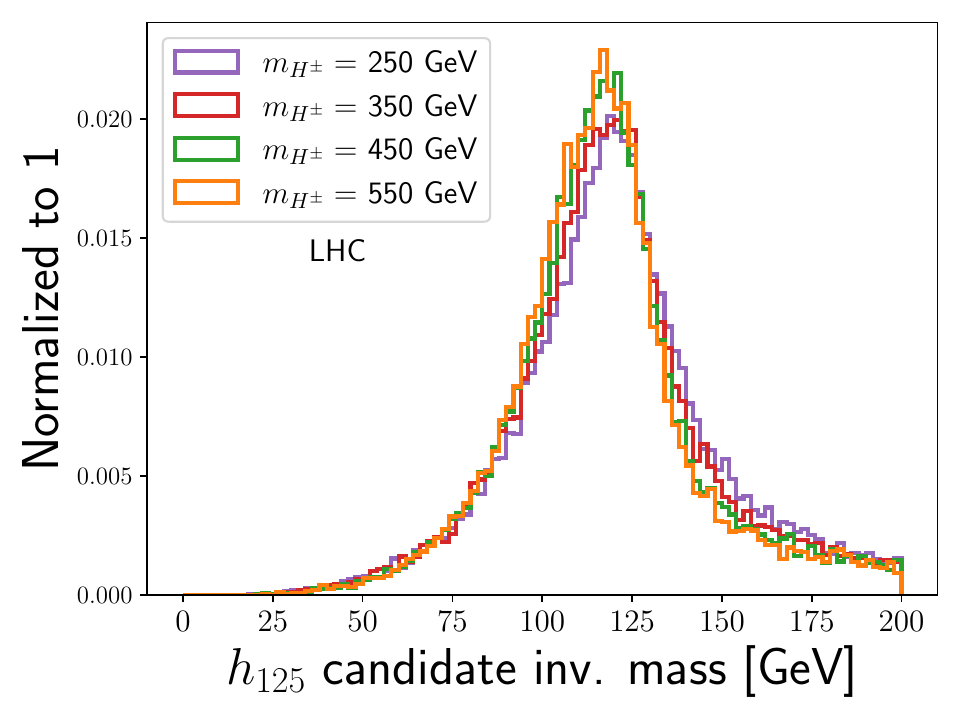}
	\caption{Invariant mass distributions of the reconstructed $\h125$ bosons in signal events.}
	\label{LHCShs}
\end{figure}

For neutral Higgs boson $\h125$ reconstruction, the method of $\chi$ minimization is applied on $b$-jet pairs with invariant masses $m_{ij}$ and $m_{kl}$:
\begin{equation}
	\chi=|m_{ij}-\mh|+|m_{kl}-\mh|.
\label{chih}
\end{equation}
The mass window for the reconstructed Higgs bosons is set to $90<\mh^{\tn{rec.}}<140$ GeV.
Figure \ref{LHCShs} shows the invariant mass distributions of the reconstructed $\h125$ pairs in signal events. For each event there is a pair of reconstructed $\h125$ whose masses are all stored and shown in Fig. \ref{LHCShs}. 

The jet four-momentum correction is also applied on light and $b$-jets to correct their pair invariant masses to the expected $W$ or $\h125$ boson masses. The correction is applied as a common factor to all four-momentum components of a given jet. If, for example, $j_1,j_2$ jets are decay products of $W \to j_1j_2$, the correction takes the following form:
\begin{align}
	p^{\mu}_{j_1}\nonumber&\to p^{\mu}_{j_1}*\frac{m_{W}}{m_{j_1j_2}},\\
	p^{\mu}_{j_2}&\to p^{\mu}_{j_2}*\frac{m_{W}}{m_{j_1j_2}}
\end{align}
with correction factors set to $m_W/m_{j_1j_2}$. 

The four-momenta of the $W$ and $\h125$ pairs are then obtained by adding four momentum components of their decay products. Therefore the final state consists of four reconstructed objects: ($W_1$,$W_2$) and ($h_1$,$h_2$) bosons which are used to find the correct pairings for the charged Higgs reconstruction. Here, $h$ is 2HDM light Higgs whose mass subscript has been temporarily dropped.  

The charged Higgs reconstruction is performed by comparing two possible pairings: $H^{\pm}_1 \to W_1 h_1$, $H^{\pm}_2 \to W_2 h_2$ vs $H^{\pm}_1 \to W_1 h_2$, $H^{\pm}_2 \to W_2 h_1$. These are the only possibilities for final state pairing. The scenario which gives closest charged Higgs invariant masses with min$|m_{\hpm_1}-m_{\hpm_2}|$ is selected and both invariant masses are stored in the final histogram.  

\subsection{Selection efficiencies}
The selection efficiencies are listed in Tab. \ref{LHCSselleff} for signal events with different $\mhpm$ and in Tab. \ref{LHCBselleff} for background samples. These efficiencies are multiplied by the signal/background cross sections and decay branching ratios and the integrated luminosity ($3000~fb^{-1}$) to get the final number of expected events at HL-LHC. 
\begin{table*}[t]
	\centering
	\begin{tabular}{|c|c|c|c|c|}
		\hline
		\multicolumn{5}{|c|}{LHC, Signal}\\
		\hline
		$\mhpm$ [GeV]&250&350&450&550\\
		\hline
		$\sigma*$BR [ab] &148.8& 89.8 & 37.4 & 17.2\\
		\hline
		2 light-jets & 135.3(0.91) & 83.8(0.93) & 35.0(0.94) & 16.1(0.93) \\
		\hline
		4 b-jets & 26.4(0.20) & 17.9(0.21) & 7.3(0.21) & 3.1(0.19) \\
		\hline
		1 lepton&11.4(0.43) & 10.2(0.57) & 4.6(0.62) & 2.0(0.65) \\
		\hline
		MET & 9.4(0.82) & 8.9(0.88) & 4.2(0.91) & 1.9(0.93) \\
		\hline
		$W$ mass window & 1.8(0.19) & 1.5(0.17) & 0.6(0.15) & 0.2(0.13) \\
		\hline
		$\h125$ mass window & 0.8(0.42) & 0.7(0.43) & 0.3(0.45) & 0.1(0.44) \\
		\hline
		Total efficiency & 0.005 & 0.007&0.007&0.006\\
		\hline
	\end{tabular}
	\caption{The signal cross section times branching ratios and the remaining values after each selection cut. Numbers in parentheses show relative efficiencies with respect to the previous cut. Total selection efficiency is shown on the last rows for the signal.}
\label{LHCSselleff}
\end{table*}

\begin{table*}[t]
	\centering
	\begin{tabular}{|c|c|c|c|c|}
		\hline
		\multicolumn{5}{|c|}{LHC, Background}\\
		\hline
		Process & $t\bar{t}$ &$t\bar{t}b\bar{b}$ & single top & VV\\
		\hline
		$\sigma*$BR [fb] & 248934 & 1406 & 262647 & 137277\\
		\hline
		2 light-jets&		227725(0.91) & 1349(0.96) & 216548(0.82) & 107090(0.78) \\
		\hline
		4 b-jet:&
		689(0.003) & 155(0.115) & 233(0.001) & 8.7(0.0001) \\
		\hline		
		1 lepton& 
		305(0.44) & 74(0.48) & 14(0.06) & 0.18(0.02) \\
		\hline		
		MET&
		250(0.82) & 62(0.83) & 11(0.83) & 0.13(0.69) \\
		\hline
		$W$ mass window&
		42.4(0.17) & 10.6(0.17) & 1.59(0.14) & 0.011(0.09) \\
		\hline
		$\h125$ mass window&
		8.27(0.19) & 2.05(0.19) & 0.29(0.18) & 0.003(0.25) \\
		\hline
		Total efficiency & 3.3$\times 10^{-5}$ & 0.001 & 1.1$\times 10^{-6}$ & 2$\times 10^{-8}$ \\
		\hline 
	\end{tabular}
	\caption{The background cross section times branching ratios and the remaining values after each selection cut. Numbers in parentheses show relative efficiencies with respect to the previous cut. Total selection efficiency is shown on the last rows for the background.}
\label{LHCBselleff}
\end{table*}
\section{Results}
\begin{figure}[hbt!]
	\centering
	\includegraphics[width=0.8\linewidth,height=0.7\linewidth]{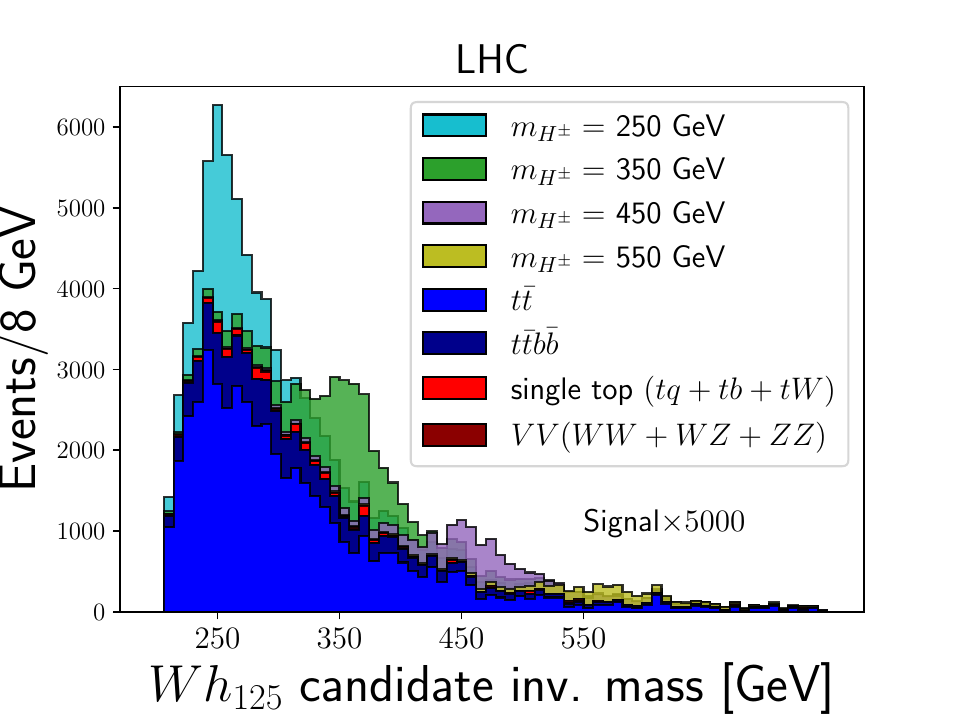}
	\caption{Invariant mass distributions of the reconstructed charged Higgs candidates on top of the SM background processes.}
	\label{LHCSB_Type1}
\end{figure}

Figure \ref{LHCSB_Type1} shows the charged Higgs invariant mass distributions on top of the SM background processes $t\bar{t}$, $t\bar{t}b\bar{b}$, single top and vector boson pair production in 2HDM type I. The signal distributions are magnified by $5000$ and are laid independently on the background. The $\alpha$ and $\beta$ parameters are fixed through $\tan\beta=10$ and $\cba=-0.2$. All distributions are normalized to integrated luminosity $\mathcal{L}=3000~fb^{-1}$ for HL-LHC. 

The signal significance is defined as $N_S/\sqrt{N_B}$ where $N_S(N_B)$ is the signal(background) number of events passed all selection cuts including a charged Higgs invariant mass window cut applied at the end. The mass window is optimized so as to give the maximum signal significance. The signal to background ratio is very small inside the charged Higgs mass window. The significance values are (15, 20, 13, 9)$\times 10^{-3}$ (for the four chosen charged Higgs masses) at $3000~fb^{-1}$. Therefore, there is no hope this channel could be observed at HL-LHC. In the final plots, we show LHC search channel ($tbH^{\pm}$) extrapolation results for HL-LHC as their expected exclusion regions are beyond the reach of $H^+H^-$ search at the same conditions. 

We proceed to evaluate the signal observability at a lepton collider in the next part. 
$~$ \\
$~$ \\
$~$ \\
\PartTitle{Part II}
\subsection{Search strategy and methodology at LC}
The full chain of the signal under consideration at LC is $e^-e^+ \to \gamma/Z/h_{125}/H \to H^+H^-$ followed by $H^+ \to W^+h_{125}$, $W\to jj$ and $h_{125}\to bb$. The contribution from neutral Higgs boson mediators is negligible compared to the electroweak contribution from $\gamma/Z$ because of the neutral Higgs small coupling with electron/positron. Therefore, the signal cross section is effectively a function of electroweak parameters and kinematics through the phase space integral. 
The main SM background processes are $t\bar{t}$ and $t\bar{t}b\bar{b}$ whose cross sections together with the signal with different charged Higgs masses are listed in Tab. \ref{xsecs}. Other SM background processes $Z/\gamma$, $ZZ$, $W^+W^-$ and $b\bar{b}b\bar{b}$ are suppressed with jet multiplicity and $b$-tagging requirements.
\begin{table}[h]
	\centering
	\begin{tabular}{|c|c|c|c|c|}
		\hline
		\multicolumn{5}{|c|}{LC, Signal}\\
		\hline
		$\mhpm$ [GeV]&250&350&450&550\\
		\hline
		$\sigma[fb]$ & 14.2 & 8.9 & 4.8 & 1.9  \\
		\hline
		\hline
		\multicolumn{5}{|c|}{LC, Background}\\
		\hline
		Process & \multicolumn{2}{c|}{$t\bar{t}$} &\multicolumn{2}{c|}{$t\bar{t}b\bar{b}$}\\
		\hline
		$\sigma[fb]$ & \multicolumn{2}{c|}{145}& \multicolumn{2}{c|}{3.8}  \\
		\hline
	\end{tabular}
	\caption{Cross sections of the signal and background samples at LC at $\sqrt{s}=1400$ GeV.}
	\label{xsecs}
\end{table}

\begin{figure}[hbt!]
	\centering
	\includegraphics[width=\linewidth,height=\linewidth]{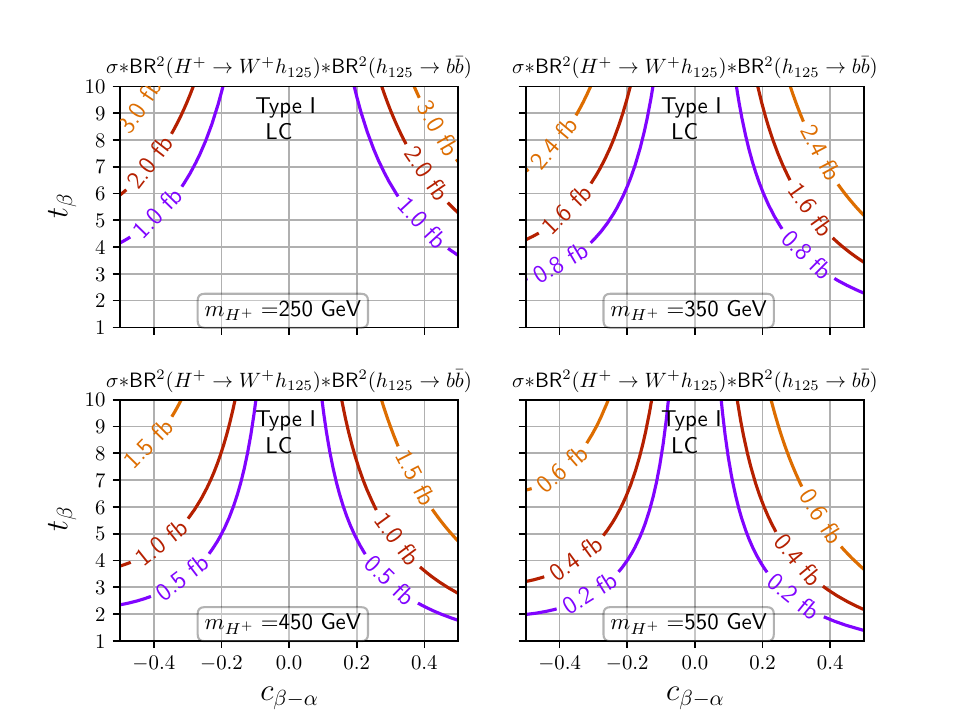}
	\caption{The product of the signal cross section times branching ratios of the charged and neutral Higgs decays in 2HDM type I as a function of $\tb,~\cba$. Results are shown for four masses $\mhpm=250,~350,~450,~550$ GeV.}
	\label{brchh}
\end{figure}

The product of the signal cross section times branching ratios of the charged and neutral Higgs decays is shown in Fig \ref{brchh} for four chosen charged Higgs masses in 2HDM type I. Plots of Fig. \ref{brchh} are the basis for final evaluation of the signal statistical significance for the chosen charged Higgs boson masses. 

The signal final state contains two light jets from $W$ and two $b-$jets from $h_{125}$. We take the advantage of larger branching ratio of hadronic decays at the expense of having multi-jet final state events. It was shown in a previous work that this type of eight jet final state can well be reconstructed if jet four-momenta are corrected and kinematic constraints are applied \cite{HR}. 
The signal and background events are generated in their hard scattering using \texttt{WHIZARD} including linear collider beam spectrum for CLIC at 1400 GeV. These events feed \texttt{PYTHIA} for particle shower generation. The full event proceeds to detector material using \texttt{DELPHES} detector simulation based on parameters defined in detector card \texttt{CLICdet-Stage2}. 

The main challenge in signal analysis in the fully hadronic final state is finding the true combination of final state jets for correct event reconstruction. We benefit from the lower particle multiplicity and hadronic activity in lepton collisions compared to LHC.

The \texttt{Valencia} jet reconstruction algorithm \cite{VLC1,VLC2} is used in the inclusive mode with the following kinematic acceptance 
\begin{equation}
	p_T^{\tn{jet}}>5 ~\GeV, ~~|\eta^{\tn{jet}}|<5
\end{equation} 
applied on the jet transverse momentum and pseudo-rapidity. The performance of the exclusive mode of the algorithm with requiring 8 jets in the final state is not better than running in the inclusive mode and then requiring at least 8 jets composed of four light and four $b$-jets. The \texttt{Valencia} algorithm has been proposed as a new clustering jet reconstruction algorithm for future $e^+e^-$ colliders whose robustness against background competes with longitudinally invariant $k_t$ algorithm used at hadron colliders \cite{VLC1}.

The $b$-tagging scenario with 90$\%$ efficiency is used as other scenarios with lower efficiencies result in less signal statistics. The miss-tagging rate includes values for the light quarks and $c$-jets as a function of the jet energy and pseudo-rapidity. Their values change from 0.01 to 0.2 for the light jets and 0.2 to 0.8 for the $c$-jets depending on the jet energy and pseudo-rapidity.

Following studies of the hadron background from photon interactions, $\gamma\gamma \to \textnormal{hadrons}$, based on CLIC proposal \cite{overlay}, a $1\%$ and $5\%$ smearing is applied on the jet momentum with $|\eta^{\tn{jet}}|<0.76$ and $|\eta^{\tn{jet}}|>0.76$ to mimic the overlay.

\begin{figure}[hbt!]
	\centering
	\includegraphics[width=0.6\linewidth,height=.6\linewidth]{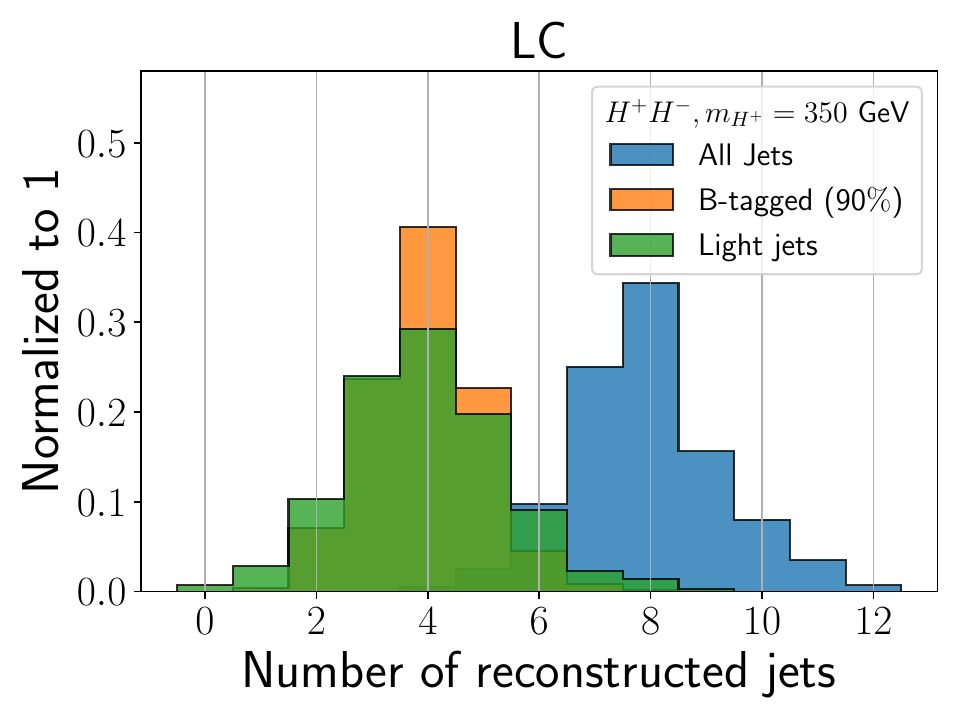}
	\caption{Number of $b$-jets, light jets and total number of jets in signal events with $\mhpm=350$ GeV.}
	\label{SJetMul}
\end{figure}

\begin{figure}[hbt!]
	\centering
	\begin{subfigure}{.23\textwidth}
		\includegraphics[width=\linewidth,height=\linewidth]{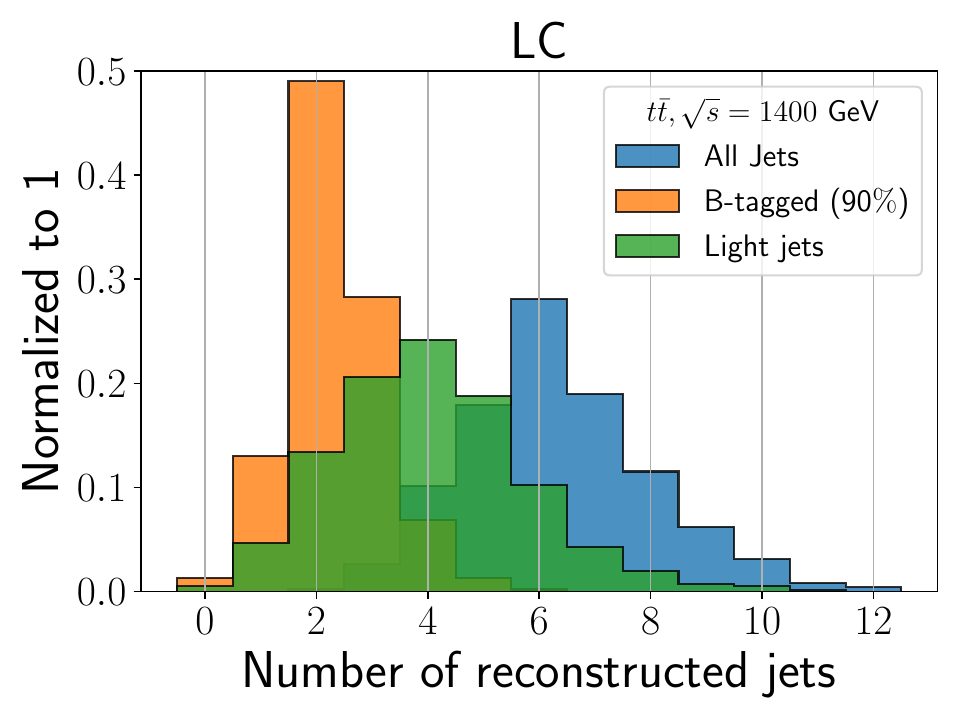}
		\caption{}
		\label{ttJetMul}
	\end{subfigure}%
	\begin{subfigure}{.23\textwidth}
		\includegraphics[width=\linewidth,height=\linewidth]{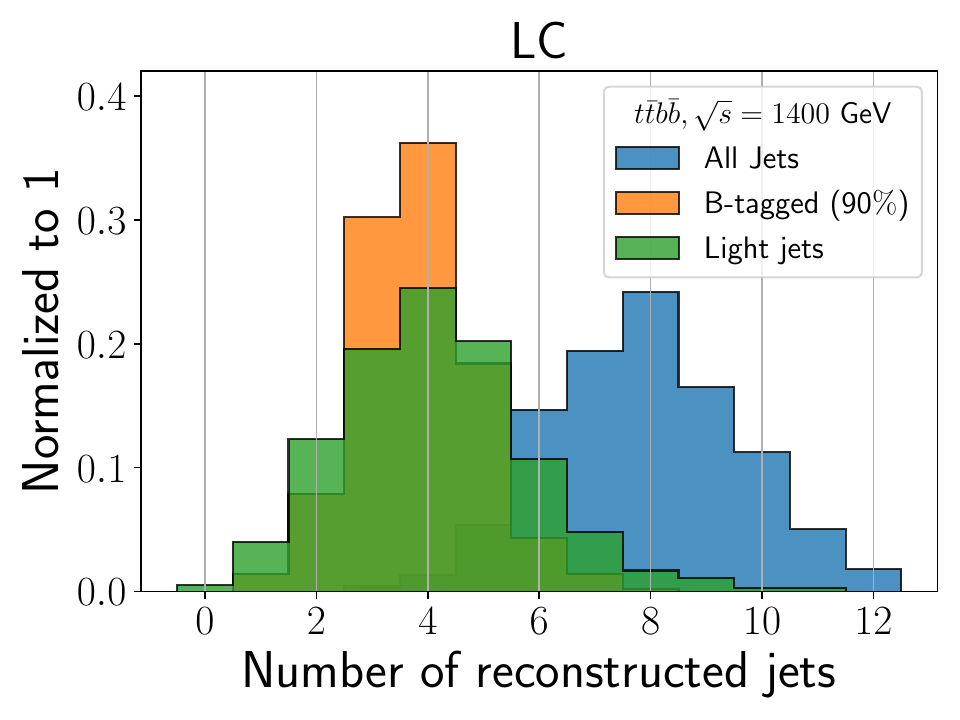}
		\caption{}
		\label{ttbbJetMul}
	\end{subfigure}
	\caption{Number of $b$-jets, light jets and total number of jets in $t\bar{t}$ (a) and $t\bar{t}b\bar{b}$ events (b).}
\end{figure}

The analysis starts with requiring at least four light jets and four $b$-jets in each event. The jet multiplicity distribution for the signal ($\mhpm=350$ GeV) is shown in Fig. \ref{SJetMul}. The corresponding distributions for $t\bar{t}$ and $t\bar{t}b\bar{b}$ are shown in Figs. \ref{ttJetMul} and \ref{ttbbJetMul} respectively. 
As seen in Fig. \ref{SJetMul} there are equal average number of four light and four $b$-jets in signal events leading to a total number of eight jets as expected. The corresponding distributions in $t\bar{t}$ and $t\bar{t}b\bar{b}$ events follow their event structure leading to an average number of two $b$-jets from the top quark decays in both background processes with an additional pair of $b$-jets present in $t\bar{t}b\bar{b}$. The number of light jets is also expected from $W\to jj$ decays.

In the next step, the method of $\chi$ minimization is applied on the jet combinations to find $W$ and $\h125$ bosons. In order to do so, we first search for two pairs of light jets $(\tn{jet}_i,\tn{jet}_j)$ and $(\tn{jet}_k,\tn{jet}_l)$ whose invariant masses $m_{ij}$ and $m_{kl}$ minimize the following $\chi$:
\begin{equation}
	\chi=|m_{ij}-m_W|+|m_{kl}-m_W|.
	\label{chiW}
\end{equation}
The mass window for the reconstructed $W$ bosons is set to $60<m_W^{\tn{rec.}}<100$ GeV.

\begin{figure}[hbt!]
	\centering
	\includegraphics[width=0.8\linewidth,height=0.7\linewidth]{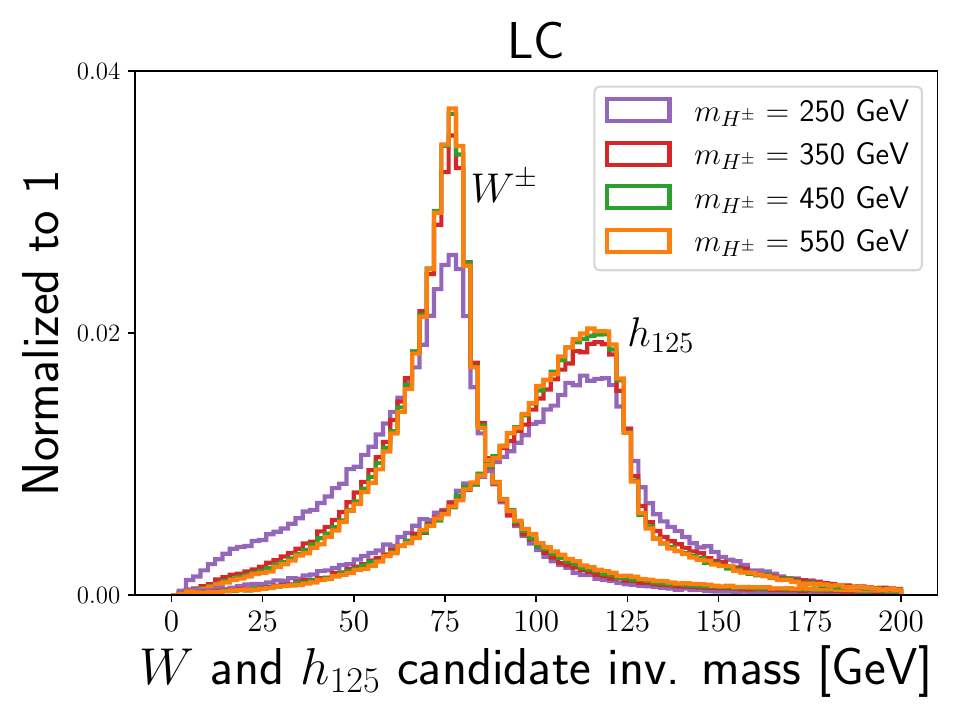}
	\caption{Invariant mass distributions of the reconstructed $W$ and $\h125$ bosons in signal events.}
	\label{SAll}
\end{figure}

The same method of $\chi$ minimization is applied on $b$-jet pairs with invariant masses $m_{ij}$ and $m_{kl}$ for neutral Higgs boson $\h125$ reconstruction:
\begin{equation}
	\chi=|m_{ij}-\mh|+|m_{kl}-\mh|.
	\label{chih}
\end{equation}
The mass window for the reconstructed Higgs bosons is set to $90<\mh^{\tn{rec.}}<140$ GeV.
Figure \ref{SAll} shows the invariant mass distributions of the reconstructed $W$ and $\h125$ pairs in signal events with different $\mhpm$ masses. For each event there is a pair of reconstructed $W$ bosons and a pair of $\h125$ whose masses are all stored and shown in Fig. \ref{SAll}. 
The jet four-momentum correction is also applied on light and $b$-jets to correct their pair invariant masses to the expected $W$ or $\h125$ boson masses as described in the previous part. 

The four-momentum components of corrected light and $b$-jet pairs are added together to obtain the $W$ and $\h125$ four-momentum vectors. Therefore the final state again consists of four reconstructed objects, i.e, two $W$ and two $\h125$ bosons which are sorted in descending energies. The main difference is that here both $W$ bosons are reconstructed in their hadronic decays. 

At this stage a kinematic selection of the final state objects is performed for the charged Higgs boson reconstruction. 

The idea is based on four-momentum conservation in the center of mass system of the collision which implies $\vec{p}_{H^{+}}=-\vec{p}_{H^{-}}$ and ${E}_{H^{+}}={E}_{H^{-}}=\sqrt{s}/2$. Suppose $H^+ \to o_1,o_4$ and $H^- \to o_2,o_3$ where $o_i$, $i=1$ to 4, denote the final state objects. The energy conservation then takes the form $E_{o_1}+E_{o_4}=E_{o_2}+E_{o_3}=\sqrt{s}/2$. If $o_1$ has the highest energy among the four, $o_4$ should have the lowest energy as their sum is fixed at $\sqrt{s}/2$. The invariant mass of the charged Higgs boson candidate is thus calculated as $m_{o_1,o_4}$ and $m_{o_2,o_3}$ after sorting final state objects in terms of their energies.

This conclusion has two sources of errors. The first is that in general beam energies are not necessarily equal due to the beam spectrum and therefore the charged Higgs energies in such cases are not equal. The second issue is about different masses of final state objects, i.e., $W$ and $\h125$ bosons. However, it is shown that the invariant mass distributions, made from $(o_1,o_4)$ and $(o_2,o_3)$ pairs, show the charged Higgs boson signal well.

\subsection{Selection efficiencies}
The selection efficiencies are listed in Tab. \ref{selleff} for signal events with different $\mhpm$ and background samples. These efficiencies are multiplied by the signal/background cross sections and decay branching ratios and the integrated luminosity ($1~ab^{-1}$) to get the final number of expected events. 
\begin{table*}[t]
	\centering
	\begin{tabular}{|c|c|c|c|c|}
		\hline
		\multicolumn{5}{|c|}{LC, Signal}\\
		\hline
		$\mhpm$ [GeV]&250&350&450&550\\
		\hline
		$\sigma*$BR [fb]&0.495& 0.772& 0.555& 0.253\\
		\hline
		4 light-jets&0.183(0.37)& 0.455(0.59)& 0.344(0.62)& 0.149(0.59)\\
		\hline
		4 $b$-jets&0.103(0.56)& 0.264(0.58)& 0.200(0.58)& 0.084(0.56)\\
		\hline
		$W$ mass window&0.036(0.35)& 0.122(0.46)& 0.092(0.46)& 0.038(0.46)\\
		\hline
		$\h125$ mass window&0.013(0.36)& 0.049(0.40)& 0.039(0.42)& 0.016(0.42)\\
		\hline
		Total efficiency&0.026& 0.063& 0.069& 0.065\\
		\hline
		\hline
		\multicolumn{5}{|c|}{LC, Background}\\
		\hline
		Process & \multicolumn{2}{c|}{$t\bar{t}$} &\multicolumn{2}{c|}{$t\bar{t}b\bar{b}$}\\
		\hline
		$\sigma*$BR [fb] &\multicolumn{2}{c|}{67.048}& \multicolumn{2}{c|}{1.757}\\ 
		\hline
		4 light-jets&\multicolumn{2}{c|}{43.581(0.65)}& \multicolumn{2}{c|}{1.107(0.63)}\\ 
		\hline
		4 $b$-jets&\multicolumn{2}{c|}{2.397(0.06)}& \multicolumn{2}{c|}{0.609(0.55)}\\ 
		\hline
		$W$ mass window&\multicolumn{2}{c|}{0.863(0.36)}& \multicolumn{2}{c|}{0.286(0.47)}\\ 
		\hline
		$\h125$ mass window&\multicolumn{2}{c|}{0.095(0.11)}& \multicolumn{2}{c|}{0.060(0.21)}\\ 
		\hline
		Total efficiency& \multicolumn{2}{c|}{0.0014}& \multicolumn{2}{c|}{0.034}  \\
		\hline
	\end{tabular}
	\caption{The signal and background cross section times branching ratios and the remaining values after each selection cut. Numbers in parentheses show relative efficiencies with respect to the previous cut. The total selection efficiency is shown on the last rows for signal and background samples. Selection efficiencies of the signal and background events.}
	\label{selleff}
\end{table*}

\section{Results}
\begin{figure}
	\centering
	\includegraphics[width=0.8\linewidth,height=0.7\linewidth]{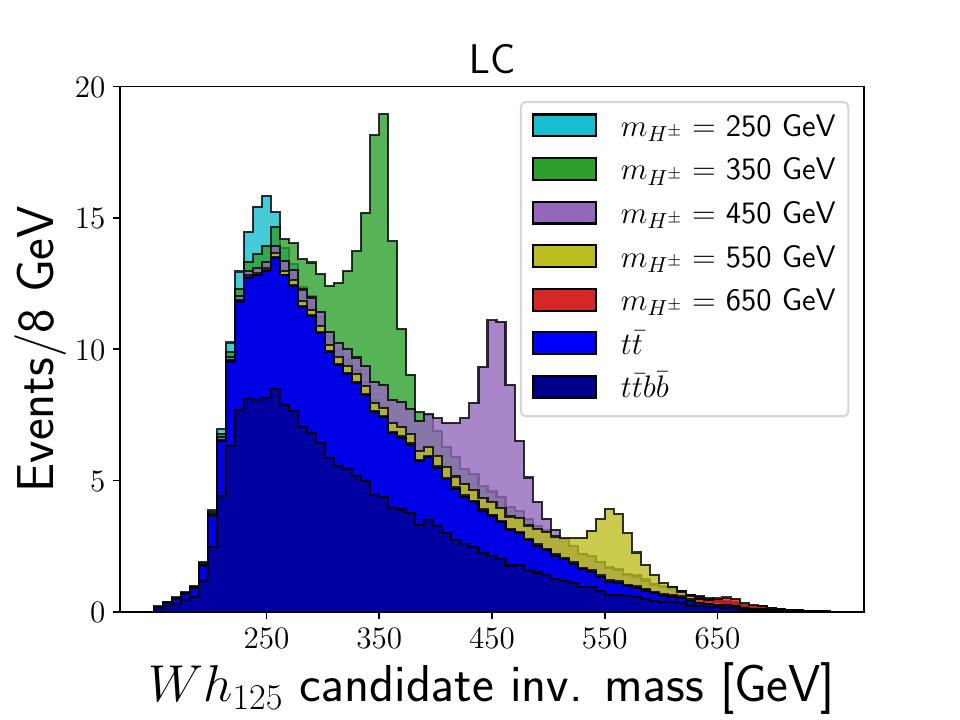}
	\caption{Invariant mass distributions of the reconstructed charged Higgs candidates on top of the SM background processes.}
	\label{SB_Type1}
\end{figure}

Figure \ref{SB_Type1} shows the charged Higgs invariant mass distributions on top of the SM background processes $t\bar{t}$ and $t\bar{t}b\bar{b}$ in 2HDM type I. The signal distributions are laid independently on the background. The $\alpha$ and $\beta$ parameters are fixed through $\tan\beta=10$ and $\cba=-0.2$. All distributions are normalized to integrated luminosity $\mathcal{L}=1~ab^{-1}$. The $b\bar{b}$ pair in $t\bar{t}b\bar{b}$ background includes sources from $Z \to b\bar{b}$ and $\h125 \to b\bar{b}$.  

The signal significance is defined as $N_S/\sqrt{N_B}$ where $N_S(N_B)$ is the signal(background) number of events passed all selection cuts. A mass window optimization is also performed to obtain the maximum signal significance as listed in Tab. \ref{ss}. 
\begin{table}[h]
	\centering
	\begin{tabular}{|c|c|c|c|c|c|c|c|c|}
		\hline
		$\mhpm$[GeV]&\multicolumn{2}{c|}{250}&\multicolumn{2}{c|}{350}&\multicolumn{2}{c|}{450}&\multicolumn{2}{c|}{550}\\
		\hline
		Mass window & 198&294&326&374&430&486&534&598\\
		\hline
$N_S$&\multicolumn{2}{c|}{15}&\multicolumn{2}{c|}{45}&\multicolumn{2}{c|}{36}&\multicolumn{2}{c|}{14}\\
		\hline
$N_B$&\multicolumn{2}{c|}{132}&\multicolumn{2}{c|}{45}&\multicolumn{2}{c|}{24}&\multicolumn{2}{c|}{8.6}\\
\hline
		Significance&\multicolumn{2}{c|}{1.3}&\multicolumn{2}{c|}{6.7}&\multicolumn{2}{c|}{7.3}&\multicolumn{2}{c|}{4.8}\\
\hline
\end{tabular}
\caption{The optimized mass window position, the number of signal and background events inside the mass window and the signal significance. Results are shown for $\tan\beta=10$ and $\cba=-0.2$ and integrated luminosity $\mathcal{L}=1~ab^{-1}$.}
\label{ss}
\end{table}
Figure \ref{SB_Type1} and Tab. \ref{ss} show results of the signal selection at a given point in parameter space. For a given charged Higgs mass, these results can be mapped to parameter space of $\cba$ vs. $\tb$ as these parameters do not alter the event kinematics and selection efficiencies. 

\begin{figure}[hbt!]
	\centering
	\includegraphics[width=\linewidth,height=\linewidth]{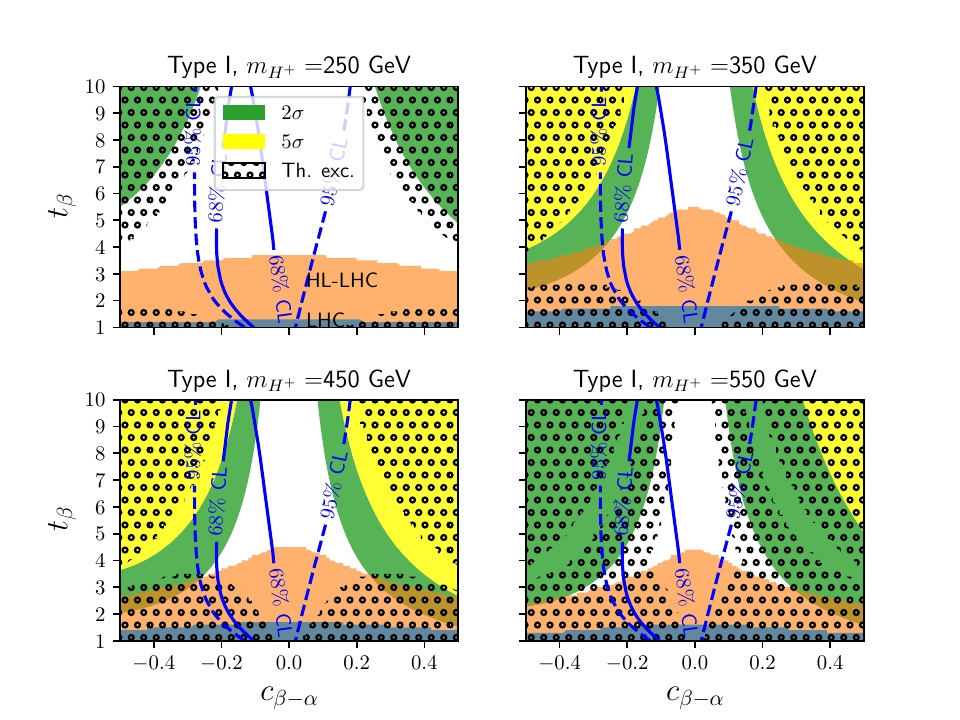}
	\caption{The signal significance in $\cba, ~\tb$ parameter space for the chosen $\mhpm$ values with $\call=1~ab^{-1}$. Contours of $2\sigma$ and $5\sigma$ significances are shown in green and yellow. Regions with different hatch styles are theoretically excluded by stability, unitarity and perturbativity requirements. The excluded region at 95$\%$ CL by LHC and also HL-LHC expectation are shown in blue and orange.}
	\label{sigma}
\end{figure}

Results shown in Fig. \ref{sigma} include $2\sigma$ and $5\sigma$ contours and the excluded regions from theoretical requirements of stability, unitarity and perturbativity. The LHC excluded region at $95\%$ CL and HL-LHC expectation are also shown in blue and orange. The solid and dashed lines show $68\%$ and $95\%$ CL contours around the best fit point with minimum $\chi^2$. 

The current LHC excluded area is limited to very low $\tb$ values. However, HL-LHC is expected to cover $\tb$ up to 5 near the alignment limit. The region of parameter space within $68\%$ CL of the best fit point is either excluded by theoretical requirements or expected to be covered by HL-LHC searches. The area inside the dashed line (95$\%$ CL around the best fit point) is not accessible for $\mhpm<250$ GeV, while $\mhpm>550$ GeV region is mostly excluded by theoretical requirements. In the mass range chosen for the analysis, the observation is that parts of the parameter space for $350<\mhpm<450$ GeV will be out of HL-LHC reach and accessible for a lepton collider to analyze.   

      \pagebreak 
\section{Conclusions}
The misaligned 2HDM type I was used as a theoretical framework for the charged Higgs search via pair production at (HL-)LHC and a lepton collider (CLIC). The charged Higgs decay $\hp \to W^+\h125$ was analyzed at LHC in semi-leptonic final state leading to no observable signal even at high luminosity era of 3000 $fb^{-1}$. 

The same channel was then analyzed at CLIC stage 2 with $\sqrt{s}=1400$ GeV including beam spectrum and effects from hadron overlay $\gamma\gamma \to \tn{hadrons}$. Fast detector simulation was performed including parametrized algorithms of $b$-tagging and its fake rate, jet reconstruction and momentum smearing for both analyses. 

The fully hadronic final state of the signal was selected at CLIC assuming $\h125\to b\bar{b}$ and $W \to jj$. Signals of the charged Higgs masses in the range $250<\mhpm<550$ GeV were analyzed assuming degenerate masses for heavy Higgs bosons $\mH=\ma=\mhpm$.

It was shown that some parts of the parameter space away from the alignment limit which are \\
$\bullet$ within 95$\%$ CL around the best fit point,\\
$\bullet$ allowed in view of theoretical requirements and\\
$\bullet$ out of HL-LHC reach,\\
can be probed by a lepton collider with signal statistical significance exceeding $2\sigma$ and $5\sigma$. The reach of such collider is limited to moderate charged Higgs masses in the range $350<\mhpm<450$ GeV and $\tb\gtrsim4$ while lower and higher masses have little chances due to small signal cross section and signal to background ratio or theoretical exclusions.

%

\end{document}